\documentclass[a4paper,11pt]{article}
\pdfoutput=1 

\usepackage{jheppub} 

\usepackage[T1]{fontenc} 
\usepackage{lmodern}
\usepackage{cancel}
\usepackage{arydshln}
\usepackage{dsfont}
\usepackage{subcaption}
\usepackage{caption}
\usepackage{xcolor}
\title{\boldmath Improved $\pi^0,\eta,\eta^{\prime}$ transition form factors in resonance chiral theory and their $a_\mu^{\rm{HLbL}}$ contribution}

\author[a]{Emilio J. Estrada,}
\author[b,c]{Sergi Gonz\`{a}lez-Sol\'is,}
\author[d]{Adolfo Guevara,}
\author[a,e]{and Pablo Roig}
\affiliation[a]{Departamento de F\'isica, Centro de Investigaci\'{o}n y de Estudios Avanzados del Instituto Polit\'{e}cnico Nacional, Apdo. Postal 14-740, 07000 Ciudad de M\'{e}xico, M\'{e}xico}
\affiliation[b]{Departament de F\'isica Qu\`{a}ntica i Astrof\'isica, Universitat de Barcelona,\\Mart\'i i Franqu\`{e}s, 1, 08028 Barcelona, Spain}
\affiliation[c]{Institut de Ci\`{e}ncies del Cosmos, Universitat de Barcelona,\\Mart\'i i Franqu\`{e}s, 1, 08028 Barcelona, Spain}
\affiliation[d]{\'{A}rea Acad\'{e}mica de Matem\'{a}ticas y F\'{i}sica, Universidad Aut\'{o}noma del Estado de Hidalgo, Carretera Pachuca-Tulancingo Km. 4.5, C.P. 42184, Pachuca, Hgo., M\'{e}xico}
\affiliation[e]{IFIC, Universitat de Val\`{e}ncia – CSIC, Catedr\'atico Jos\'e Beltr\'an 2, E-46980 Paterna, Spain}

\emailAdd{emilio.estrada@cinvestav.mx}
\emailAdd{sergig@icc.ub.edu}
\emailAdd{adolfo\_guevara@uaeh.edu.mx}
\emailAdd{pablo.roig@cinvestav.mx}

\abstract{Working with Resonance Chiral Theory, within the two resonance multiplets saturation scheme, we satisfy leading (and some subleading) chiral and asymptotic QCD constraints and accurately fit simultaneously the $\pi^{0},\eta,\eta^{\prime}$ transition form factors, for single and double virtuality. 
In the latter case, we supplement the few available measurements with lattice data to ensure a faithful description.
Mainly due to the new results for the doubly virtual case, we improve over existing descriptions for the $\eta$ and $\eta^\prime$. 
Our evaluation of the corresponding pole contributions to the hadronic light-by-light piece of the muon $g-2$ read:
$a_\mu^{\pi^{0}\text{-}\rm{pole}}=\left(61.9\pm0.6^{+2.4}_{-1.5}\right)\times10^{-11}$, $a_\mu^{\eta\text{-}\rm{pole}}=\left(15.2\pm0.5^{+1.1}_{-0.8}\right)\times10^{-11}$ and $a_\mu^{\eta^\prime\text{-}\rm{pole}}=\left(14.2\pm0.7^{+1.4}_{-0.9}\right)\times10^{-11}$, for a total of $a_\mu^{\pi^0+\eta+\eta^{\prime}\text{-}\rm{pole}}=\left(91.3\pm1.0^{+3.0}_{-1.9}\right)\times10^{-11}$, where the first and second errors are the statistical and systematic uncertainties, respectively.}

\begin{document} 
\maketitle
\flushbottom

\section{Introduction}\label{sec_Intro}

The anomalous magnetic moment of the muon ($a_\mu=(g_\mu-2)/2$, with $g$ the gyromagnetic factor) has been one of the most attractive observables in recent years, given its extremely precise measurements~\cite{Muong-2:2006rrc,Muong-2:2021ojo, Muong-2:2023cdq} and Standard Model predictions~\cite{Aoyama:2020ynm}\footnote{The Standard Model prediction in the White Paper~\cite{Aoyama:2020ynm} is based on refs.~\cite{Davier:2017zfy,Keshavarzi:2018mgv,Colangelo:2018mtw,Hoferichter:2019mqg,Davier:2019can,Keshavarzi:2019abf,Kurz:2014wya,FermilabLattice:2017wgj,Budapest-Marseille-Wuppertal:2017okr,RBC:2018dos,Giusti:2019xct,Shintani:2019wai,FermilabLattice:2019ugu,Gerardin:2019rua,Aubin:2019usy,Giusti:2019hkz,Melnikov:2003xd,Masjuan:2017tvw,Colangelo:2017fiz,Hoferichter:2018kwz,Gerardin:2019vio,Bijnens:2019ghy,Colangelo:2019uex,Pauk:2014rta,Danilkin:2016hnh,Jegerlehner:2017gek,Knecht:2018sci,Eichmann:2019bqf,Roig:2019reh,Colangelo:2014qya,Blum:2019ugy,Aoyama:2012wk,Aoyama:2019ryr,Czarnecki:2002nt,Gnendiger:2013pva}. 
This field has been evolving very actively since ref.~\cite{Aoyama:2020ynm} was published. See {\it{e.g.}}, refs.~\cite{Colangelo:2022jxc, Lehner:2020crt,Hoferichter:2020lap,Knecht:2020xyr, Masjuan:2020jsf, Ludtke:2020moa, Chao:2020kwq, Miranda:2020wdg, Hoid:2020xjs, Bijnens:2020xnl, Colangelo:2020lcg, Bijnens:2021jqo, Chao:2021tvp, Danilkin:2021icn, Colangelo:2021nkr, Leutgeb:2021mpu, Colangelo:2021moe, Cappiello:2021vzi, Miranda:2021lhb, Giusti:2021dvd, Hoferichter:2021wyj, Miramontes:2021exi, Stamen:2022uqh, Boito:2022rkw, Wang:2022lkq, Aubin:2022hgm, Colangelo:2022vok, Ce:2022kxy, ExtendedTwistedMass:2022jpw, Colangelo:2022prz, Biloshytskyi:2022ets, Boito:2022dry, Leutgeb:2022lqw,Bijnens:2022itw, Colangelo:2023een,FermilabLatticeHPQCD:2023jof, RBC:2023pvn, Davier:2023hhn, Wang:2023njt, Ludtke:2023hvz, Escribano:2023seb, Blum:2023vlm, Masjuan:2023qsp, Nesterenko:2023bdc, Benton:2023dci, Hoferichter:2023sli, Davier:2023cyp, Davier:2023fpl,Colangelo:2024xfh,Hoferichter:2024fsj, Boito:2024yat,Boccaletti:2024guq,DiLuzio:2024sps,Lahert:2024vvu,Keshavarzi:2024wow,Budassi:2024whw}.} and the persistent discrepancy between them, which seems to be reducing lately~\cite{Masjuan:2023qsp,Davier:2023fpl,CMD-3:2023alj,CMD-3:2023rfe,Boccaletti:2024guq}.
While the experimental uncertainty on $a_\mu$ has shrunk up to $22\times10^{-11}$, and will likely be $\lesssim16\times10^{-11}$ after the final FNAL measurement, the error of the theory prediction is not at the same level at the moment.

This uncertainty is dominated by the hadronic interaction contributions in the non-perturbative regime of QCD, where hadrons play the key r\^ole. Among these, the hadronic vacuum polarization (HVP) part dominates the error, yielding $40$ of the $43$ units (in $10^{-11}$) of the overall uncertainty, the rest coming from the sum in quadrature of this error with the one stemming from the hadronic light-by-light (HLbL) piece, $17\times10^{-11}$. 
The uncertainty of the HVP part in the new evaluation of ref.~\cite{Boccaletti:2024guq} ($33\times10^{-11}$) is, however, smaller than the one in the White Paper~\cite{Aoyama:2020ynm}. 
The error in ref.~\cite{Boccaletti:2024guq} still basically doubles that of the HLbL contribution, which is commensurate with the expected uncertainty of the final FNAL $a_\mu$ publication. 
Therefore, although not pressing now, the need to improve over existing computations of $a_\mu^{\rm{\rm{HLbL}}}$ will be intensified by the time the $a_\mu^{\rm{HVP}}$ picture becomes clearer (the most recent developments seem to be pointing that way). 
In this context, it is the purpose of this work to advance in the understanding of the dominant contribution to $a_\mu^{\rm{HLbL}}$ (yielding its whole value~\cite{Aoyama:2020ynm} at $68\%$ C.L.\footnote{It must be clarified that it contributes to the $a_\mu^{\rm{HLbL}}$ uncertainty, $17\times10^{-11}$, only with $5\times10^{-11}$ directly, although its dominant part may be ascribed to the interplay of $P$ and axial poles, related to the fulfillment of QCD short-distance constraints and the chiral anomaly, as commented in the next paragraph.} $\sim93\times10^{-11}$), given by the lightest pseudoscalar poles: those corresponding to the $\pi^0$, $\eta$ and $\eta^\prime$ mesons, $a_\mu^{P\text{-}\rm{poles},\,HLbL}$.

It was soon realized that the bulk of $a_\mu^{\rm{HLbL}}$ comes from the $P$-poles contribution~\cite{deRafael:1993za,Hayakawa:1995ps, Bijnens:1995cc, Bijnens:1995xf, Hayakawa:1996ki, Hayakawa:1997rq, Bijnens:2001cq}. 
Noticeably, the usefulness of the limit of  QCD for a large number of colors ($N_C$)~\cite{tHooft:1973alw,tHooft:1974pnl,Witten:1979kh} and the chiral counting~\cite{Weinberg:1978kz,Gasser:1983yg,Gasser:1984gg} were acknowledged in comprehending the relative size of the various contributions to $a_\mu^{\rm{HLbL}}$ and their numerical ballpark values~\cite{deRafael:1993za,Bijnens:1995cc,Bijnens:1995xf,Bijnens:2001cq}. 
Large-$N_C$ QCD and the short-distance constraints of the strong interactions were the guiding principles in ref.~\cite{Knecht:2001qf}, which was a key source for later evaluations of $a_\mu^{P\text{-}\rm{poles},\,HLbL}$. 
Particularly, the short-distance behaviour of the form-factor entering the real photon vertex was challenged in ref.~\cite{Melnikov:2003xd} and the importance of axial-vector mesons contributions in fulfilling asymptotic QCD was put forward, a topic which has advanced notably since then (see e.~g.~refs.~\cite{Jegerlehner:2017gek,Roig:2019reh,
Hoferichter:2020lap,Knecht:2020xyr,Masjuan:2020jsf, Ludtke:2020moa, Bijnens:2020xnl, Bijnens:2021jqo,  Colangelo:2021nkr,Leutgeb:2021mpu,Miranda:2021lhb,Leutgeb:2022lqw,Bijnens:2022itw, Ludtke:2023hvz,Pauk:2014rfa,Leutgeb:2019gbz,Cappiello:2019hwh,Zanke:2021wiq, Leutgeb:2021bpo, Radzhabov:2023odj, Hoferichter:2023tgp}).

We recapitulate the situation for $a_\mu^{P\text{-}\rm{poles},\,HLbL}$ by the White Paper. 
The $\pi^0$ contribution was taken from the dispersive evaluation~\cite{Hoferichter:2018kwz}, which agreed with an earlier computation based on rational approximants~\cite{Masjuan:2017tvw}. 
The latter was the one taken for the $\eta^{(\prime)}$ contributions, for which a numerical dispersive result has not been published yet. 
Dyson-Schwinger evaluations~\cite{Eichmann:2019tjk,Raya:2019dnh} and the lattice QCD computation~\cite{Gerardin:2019vio} (including only the $\pi^0$) agreed with them within errors. 
These results are summarized in the first rows of table~\ref{tab:amuEvaluations}, yielding the data-driven prediction $a_\mu^{P\text{-}\rm{poles,\,HLbL}}=\left(93.8^{+4.0}_{-3.6}\right)\times10^{-11}$~\cite{Aoyama:2020ynm}. 
The numerical values of other data-driven calculations of $a_\mu^{P\text{-}\rm{poles,\,HLbL}}$ were not quoted in the White Paper~\cite{Aoyama:2020ynm}, because they failed to satisfy the quality criteria listed just before its section 4.4.1., which we copy here--for reference--at the beginning of section~\ref{sec_DataAnalysis}.

This was precisely the case of computations using Resonance Chiral Lagrangians~\cite{Ecker:1989yg}: refs.~\cite{Kampf:2011ty,Roig:2014uja} and~\cite{Guevara:2018rhj}, which added flavor-symmetry breaking corrections to the previous two calculations (see also ref.~\cite{Czyz:2012nq}). 
In particular, including just one multiplet of vector mesons it was not possible to comply with the leading asymptotic behaviour for double virtuality, thus violating the last part of the first requirement for the White Paper~\cite{Aoyama:2020ynm}. 
In ref.~\cite{Guevara:2018rhj}, the associated error was estimated to be $\left(^{+5.0}_{-0.0}\right)\times 10^{-11}$, from the contribution of excited vector multiplets, which would restore the appropriate short-distance behaviour for double virtuality. 
In this work we have verified previous results and computed all new contributions associated to the first resonance excitations within this formalism, being consistent with the procedure of ref.~\cite{Guevara:2018rhj} and complying now with the QCD short-distance behaviour.\footnote{A more complicated solution, including three resonance multiplets, was worked out in ref.~\cite{Kadavy:2022scu}. 
In this case it is possible to match even more subleading QCD asymptotic constraints. 
See table~\ref{tab:DiffswithPrague} for the comparison of our procedure to this reference.}

Resonance Chiral Theory (R$\chi$T)~\cite{Ecker:1989yg} is an effective approach that adds, to the Chiral Perturbation Theory ($\chi$PT)  action~\cite{Weinberg:1978kz, Gasser:1983yg}, the light-flavored resonances as active fields. 
This is done without any assumption on the resonance dynamics. 
In this way, one can for instance see the celebrated notion of vector meson dominance emerging as a result of integrating out the resonance fields and checking that the next-to-leading order chiral low-energy constants are essentially saturated by the corresponding contributions to the chiral effective Lagrangian. 
We work within the large-$N_C$ limit and assume a $1/N_C$ expansion (the small impact of subleading terms is considered in section \ref{sec_NLO1/N}). 
Thus, the spontaneous symmetry breaking pattern is $U(3)_L\times U(3)_R\to U(3)_{L+R\sim V}$, so that the $P$ pseudoscalars will be part of a nonet of pseudo-Goldstone bosons. 
We will consider the R$\chi$T Lagrangian pieces derived in refs.~\cite{Ecker:1989yg, Kampf:2011ty, Ruiz-Femenia:2003jdx, Cirigliano:2006hb, Cirigliano:2003yq, Guo:2009hi,Mateu:2007tr,Kadavy:2020hox}. 
The restrictions imposed by perturbative QCD on the asymptotic behaviour of the $P$ transition form factors will determine all but $12$ parameters (including four describing the $\eta$-$\eta^\prime$ mixing), that will be fitted to the available spacelike data: the di-photon $P$ decay widths, and the singly and doubly virtual measurements. 
For double virtuality, supplementing the very few points (only for the $\eta^\prime$) with lattice data will turn out to be fundamental.

As the main outcome of our work, we predict the following $a_\mu^{P\text{-}\rm{poles,\,HLbL}}$ contributions (in units of $10^{-11}$): $a_\mu^{\pi^{0}\text{-}\rm{pole}}=\left(61.9\pm0.6^{+2.4}_{-1.5}\right)$, $a_\mu^{\eta\text{-}\rm{pole}}=\left(15.2\pm0.5^{+1.1}_{-0.8}\right)$ and $a_\mu^{\eta^\prime\text{-}\rm{pole}}=\left(14.2\pm0.7^{+1.4}_{-0.9}\right)$, for a total of $a_\mu^{\pi^0+\eta+\eta^{\prime}\text{-}\rm{pole}}=\left(91.3\pm1.0^{+3.0}_{-1.9}\right)$, where the first uncertainty is statistical and the second one is the systematic theory error (discussed in detail in section~\ref{sec_TheoryErrors}).

The paper is organized as follows. In section \ref{sec_Lagrangian}, we review the relevant pieces of the resonance chiral Lagrangian. After that, in section \ref{sec_TFFs}, we introduce briefly the $P
$ transition form factors and focus on the new contributions, from the excited resonances, that were disregarded in ref.~\cite{Guevara:2018rhj}. We start section \ref{sec_SDCs} recalling the restrictions imposed by perturbative QCD on the asymptotic behaviour of the $P$ transition form factors. Then, we apply them to our results, first in the chiral limit and then including $\mathcal{O}(m_P^2)$ corrections. 
The relations that we found on the resonance Lagrangian couplings are consistent with the restrictions obtained studying the $VVP$ Green's function. 
Using these relations allows us to get more compact form factors, depending on 12 parameters, that are confronted to spacelike data (for real photons, and for the singly- and doubly-virtual cases) in section~\ref{sec_DataAnalysis}. 
We quote and discuss our best fit results, verifying their accuracy and consistency. 
Section~\ref{sec_CA} examines their correspondence with those obtained with Canterbury Approximants, explaining our preference for R$\chi$T. We compute the $P$-pole contributions to $a_\mu$ in section~\ref{sec_amuPpole} and state our conclusions in section~\ref{sec_Concl}.

\section{Resonance Chiral Theory Lagrangian}\label{sec_Lagrangian}

In this section we describe the formalism used to compute the $P$ transition form factors, namely R$\chi$T,\footnote{A more extended discussion can be found in e.g.~ref.~\cite{Guevara:2018rhj}.} which extends the energy domain of applicability of the $\chi$PT Lagrangian~\cite{Gasser:1983yg} by including the light-flavored resonances as explicit degrees of freedom~\cite{Ecker:1989yg}.

The leading contribution to the $P\gamma\gamma$ transition form factor (TFF) at low energies\footnote{It is $\mathcal{O}(p^4)$ in the chiral expansion, where $p^2\sim m_P^2$~\cite{Gasser:1983yg}.} is given by the Wess-Zumino-Witten (WZW) contact term~\cite{Wess:1971yu, Witten:1983tw}, which is completely specified by the chiral anomaly~\cite{Adler:1969gk,Bell:1969ts}, in terms of the number of colours of the QCD gauge group ($N_C$) and the pion decay constant in the chiral limit ($F$). 
We will not quote here the WZW action (it can be found in e.g. appendix A of ref.~\cite{Guevara:2018rhj}), but rather recall in the next section its contribution to the $\pi^{0},\eta$ and $\eta^{\prime}$ TFF. 
This one, being a constant, demands additional pieces to yield a behaviour complying with QCD constraints. 
Within R$\chi$T we will accomplish this by considering interactions mediated by vector mesons that will modify the WZW local interaction. 
We will need to account for odd-intrinsic parity operators with a pseudo-Goldsonte boson and either two vector resonances, or one such resonance and a vector current. We will also need to include the non-resonant $R\chi T$ odd-intrinsic parity Lagrangian at $\mathcal{O}(p^6)$, which has the same operator structure as the $\chi$PT Lagrangian of the same order~\cite{Bijnens:2001bb}.

The complete basis of odd-intrinsic parity R$\chi$T operators which -upon resonance integration- saturate most of the $\mathcal{O}(p^6)$ chiral low-energy constants (LECs) was found in ref.~\cite{Kampf:2011ty}. We will, however, use the basis in ref.~\cite{Ruiz-Femenia:2003jdx}, which is sufficient for processes including just one pseudoscalar meson~\cite{Roig:2013baa} and simpler (smaller) for the present study. 
To achieve a consistent description~\cite{Kampf:2011ty}, we need to consider pseudoscalar resonances as well~\cite{Guevara:2018rhj}, which naturally come along with the second vector multiplet~\cite{Roig:2014uja} that we introduce here in the context of the $P$ TFF within R$\chi$T.

We organize the Lagrangian describing the lightest pseudo-Goldstone bosons and their interactions in increasing number of resonance fields, $R$:
\begin{equation}\label{eq_LRChT}
\mathcal{L}_{\rm{R\chi T}}=\mathcal{L}_{\mathrm{no}\;R}+\sum_R (\mathcal{L}_R^{\rm{Kin}}+\mathcal{L}_{R^\prime}^{\rm{Kin}})+\sum_{R,R'} \mathcal{L}_{R,R'}+\cdots\,.
\end{equation}
We will further divide the R$\chi$T Lagrangian into its odd- and even-intrinsic parity sectors, including, in addition to the pseudo-Goldstone bosons ($\varphi^a$) and the photons ($\gamma$), the two lightest vector meson multiplets ($V^{(\prime)}$) and the first pseudoscalar excitations ($P^\prime$).\\

\textbf{$\bullet$ Operators with no resonance fields:}\\[1ex]
This part of the Lagrangian will be formally equivalent to the $\chi$PT one. 
We must note, however, that the values of the chiral LECs vary from $\chi$PT to this sector of R$\chi$T precisely by the fact that in the latter the resonance degrees of freedom are active, whereas they have been integrated out in the former. 
The relevant part of $\mathcal{L}_{{\mathrm{no}}\;R}$ in both intrinsic parity sectors reads ($\langle...\rangle$ stands for a trace in flavor space):~\footnote{We include the $\mathcal{O}_j^W$ operators, which are subleading in the chiral expansion with respect to the WZW piece, to ensure the most general breaking of flavor symmetry. Nevertheless, our analysis of short-distance constraints will show that their coefficients, $C_j^W$, vanish.}
\begin{equation}\label{eq.L_noRes}
        \mathcal{L}_{\mathrm{no}\;R}^{\mathrm{even}}=\frac{F^2}{4}\langle u_\mu u^\mu +\chi_+\rangle\,,\quad \mathcal{L}_{\mathrm{no}\;R}^{\mathrm{odd}}=\mathcal{L}_{\rm{WZW}}+\sum_{j=7,8,22}C_j^W \mathcal{O}_j^W\,,
\end{equation}
where we used the chiral tensors~\cite{Bijnens:1999sh}
\begin{equation}\label{eq_chiraltensors}
            u_\mu=i\left[u^\dagger(\partial_\mu-ir_\mu)u-u(\partial_\mu-il_\mu)u^\dagger\right],\quad 
            \chi_\pm=u^\dagger\chi u^\dagger \pm u \chi^\dagger u\,, 
\end{equation}
in which the external (pseudo)scalar ($p$)$s$ spin-zero and (left)-right ($l_\mu$)$r_\mu$ spin-one sources appear. 
For the vector electromagnetic case
\begin{equation} \label{eq_spin1sources}
        r_\mu=l_\mu=eQA_\mu+\cdots, \quad Q=\mathrm{diag}\left(\frac{2}{3},-\frac{1}{3},-\frac{1}{3}\right)\,.
\end{equation}
Explicit chiral symmetry breaking is introduced like in QCD, through non-vanishing quark masses, by means of the scalar current, via
\begin{equation}\label{eq_spin0sources}
\chi=2B
(s+ip),\quad s=\mathrm{diag}(m_u,m_d,m_s),\quad p=0, \quad 2m=m_u+m_d,
\end{equation}
where the last equality corresponds to the isospin symmetry limit\footnote{The isospin symmetry limit corresponds to $m_u=m_d$, denoted here $m$, resulting in $m_u+m_d=2m$.}, which is an excellent approximation in our study.
In eq.~(\ref{eq_chiraltensors}), the operator $u$ also appears, depending non-linearly on the pseudo-Goldstone bosons $\varphi^a$ ($\sqrt{2}\Phi=\sum_{a=0}^8 \varphi^a\lambda_a$, the $\eta^{\prime}$ becomes the ninth such boson in the chiral and large-$N_C$ limits~\cite{tHooft:1973alw,tHooft:1974pnl,Witten:1979kh})
\begin{equation}
            u= \exp\left(i\frac{\Phi}{\sqrt{2}F}\right)\,,
\end{equation}
through
\begin{equation}\label{eq.PhiMatrix}
        \Phi=\begin{pmatrix}
            \frac{1}{\sqrt{2}}(C_\pi \pi^0 +C_q\eta+C_q^{\prime} \eta^{\prime}) & \pi^+& K^+\\
            \pi^-&\frac{1}{\sqrt{2}}(-C_\pi \pi^0 +C_q\eta+C_q^{\prime} \eta^{\prime}) & K^0\\
            K^-& \Bar{K^0}& -C_s\eta+C_s^\prime \eta^{\prime}\\
        \end{pmatrix}\,.
\end{equation}
 In eq.~(\ref{eq.PhiMatrix}), $C_\pi=F/F_\pi$ in the large-$N_C$ limit~\cite{Bernard:1991zc, Sanz-Cillero:2004hed,Guo:2014yva}, with the physical pion decay constant $F_\pi=92.2$ MeV~\cite{Workman:2022ynf}.
$C_{q,s}^{(\prime)}$ describe the $\eta$-$\eta^\prime$ system in the two-angle mixing scheme
\begin{subequations}\label{eq_2mixingangles}
    \begin{equation}
        C_q=\frac{F}{\sqrt{3}\cos{(\theta_8-\theta_0})}\left(\frac{\cos{\theta_0}}{f_8}-\frac{\sqrt{2}\sin{\theta_8}}{f_0}\right)\,,
    \end{equation}
    \begin{equation}
        C_{q}^{\prime}=\frac{F}{\sqrt{3}\cos{(\theta_8-\theta_0)}}\left(\frac{\sqrt{2}\cos{\theta_8}}{f_0}+\frac{\sin{\theta_0}}{f_8}\right)\,,
    \end{equation}
    \begin{equation}
        C_s=\frac{F}{\sqrt{3}\cos{(\theta_8-\theta_0})}\left(\frac{\sqrt{2}\cos{\theta_0}}{f_8}+\frac{\sin{\theta_8}}{f_0}\right)\,,
    \end{equation}
    \begin{equation}
        C_{s}^{\prime}=\frac{F}{\sqrt{3}\cos{(\theta_8-\theta_0)}}\left(\frac{\cos{\theta_8}}{f_0}-\frac{\sqrt{2}\sin{\theta_0}}{f_8}\right)\,,
    \end{equation} 
    \label{eq.two-angle mixing}
\end{subequations}
in terms of two angles ($\theta_{8,0}$) and two couplings ($f_{8,0}$)~\cite{Schechter:1992iz, Bramon:1997va, Feldmann:1999uf, Escribano:2005qq}.

The relevant operators entering $\mathcal{L}_{\mathrm{no}\;R}$ in eq.~(\ref{eq.L_noRes}) are ($\epsilon_{\mu\nu\rho\sigma}$ is common to all of them and omitted below, we use the standard convention $\epsilon_{0123}=+1$)\footnote{The operator $\mathcal{O}_8^W$ is suppressed in the $1/N_C$ expansion since it has a double-trace. We keep it because this type of corrections could in principle play a non-negligible r\^ole for the $\eta$-$\eta^\prime$ mesons~\cite{Guevara:2018rhj}.}
\begin{equation}\label{eq_ChPTOp6ops}
\mathcal{O}_7^W=i\langle \chi_- f_+^{\mu\nu} f_+^{\rho\sigma}\rangle\,,\quad \mathcal{O}_8^W=i\langle\chi_-\rangle \langle f_+^{\mu\nu} f_+^{\rho\sigma}\rangle\,,\quad \mathcal{O}_{22}^W=\langle u^\mu \left\{ \nabla_\lambda f_+^{\lambda\nu},f_+^{\rho\sigma}\right\}\rangle\,,
\end{equation}
where $f_\pm^{\mu\nu}=uF_L^{\mu\nu}u^\dagger \pm u^\dagger F_R^{\mu\nu}u$, and $F_{L(R)}^{\mu\nu}=\partial^\mu\ell(r)^\nu-\partial^\nu\ell(r)^\mu-i[\ell(r)^\mu,\ell(r)^\nu]=eQF^{\mu\nu}+\cdots,$ in terms of the electromagnetic field-strength tensor, $F^{\mu\nu}$. 
The covariant derivative is defined as $\nabla_\mu \cdot =\partial_\mu \cdot+[\Gamma_\mu,\cdot]$, with the connection $\Gamma_\mu =\frac{1}{2}\left[ u^\dagger(\partial_\mu-ir_\mu)u+u(\partial_\mu-il_\mu)u^\dagger \right]$. 
In eq.~(\ref{eq_ChPTOp6ops}), only $\mathcal{O}_{22}$ does not induce $U(3)$ symmetry breaking corrections.\\

\textbf{$\bullet$ Vector resonances and their kinetic terms:}\\[1ex]
We describe the resonances with antisymmetric tensor fields~\cite{Ecker:1989yg}. 
In the vector case, this corresponds to
\begin{equation}\label{eq_Vmatrix}
        V_{\mu\nu}=\begin{pmatrix}
        (\rho^0+\omega^0)/\sqrt{2} & \rho^+ & K^{*+}\\
        \rho^-& (-\rho^0+\omega^0)/\sqrt{2}& K^{*0}\\
        K^{*-} & \Bar{K}^{*0} & \phi 
        \end{pmatrix}_{\mu\nu},
\end{equation}
where the ideal $\omega$-$\phi$ mixing scheme (in which $\omega=\sqrt{\frac{2}{3}}\omega_8+\frac{1}{\sqrt{3}}\omega_0$) has been used, as obtained in the large-$N_C$ limit. The kinetic terms for the $V$ resonances read
\begin{equation}\label{eq_Kin}
        \mathcal{L}_V^{\mathrm{Kin}}=-\frac{1}{2}\langle \nabla_\lambda V^{\lambda\nu}\nabla^\rho V_{\rho\nu}\rangle + \frac{1}{4}M_V^2\langle V_{\mu\nu}V^{\mu\nu}\rangle-e^V_m \langle V_{\mu\nu}V^{\mu\nu}\chi_+\rangle\,,
\end{equation}
where a common $U(3)$-symmetric mass, $M_V$, appears. It is corrected by the $e^V_m$-term, yielding a pattern according to phenomenology \cite{Cirigliano:2003yq, Guo:2009hi}. We note that interactions are hidden in the expansion of the covariant derivatives above, but these play no r\^ole in our work. We turn now to the relevant interactions involving at least one vector resonance field.\\

\textbf{$\bullet$ Operators with one vector resonance field:}\\[1ex]
We start with the even-intrinsic parity sector, where we have
\begin{equation}\label{eq.Lagrangin1Reven}
\mathcal{L}_V^{\mathrm{even}}=\frac{F_V}{2\sqrt{2}}\langle V_{\mu\nu}f_+^{\mu\nu}\rangle+\frac{\lambda_V}{\sqrt{2}}\langle V_{\mu\nu} \left\{f_+^{\mu\nu},\chi_+\right\}\rangle\,,
\end{equation}
with the first term giving the leading contribution to the $V^0$-$\gamma$ coupling and the second one including its flavor-symmetry breaking correction, proportional to quark masses.\footnote{We note that $\lambda_V=\sqrt{2}\lambda^V_6$ in ref.~\cite{Cirigliano:2006hb}.}

In the odd-intrinsic parity sector one has
\begin{equation}
\mathcal{L}_V^{\mathrm{odd}}=\sum_{j=1}^7 \frac{c_j}{M_V} \mathcal{O}_{VJP}^j\,,
\end{equation}
where the complete list of operators contributing to our processes of interest (via $\varphi$-$\gamma$-$V$ vertices) is collected in table \ref{tab:OVJPops}. 
Among these operators only $\mathcal{O}_{VJP}^3$ breaks chiral symmetry \footnote{Although $\mathcal{O}_{VJP}^4$ also breaks chiral symmetry, its contribution vanishes for the processes studied here.}(the others will also give flavor-breaking contributions for real $\varphi$s, once their wave functions renormalizations and mixings are accounted for; with similar comments applying below).\\

\begin{table}[h!]
    \centering
    \begin{tabular}{| c | c |}\hline
    $j$ & $\mathcal{O}^j_{VJP}$  \\ \hline
    1 & $\langle\{V^{\mu\nu},f_{+}^{\rho\alpha}\}\nabla_\alpha u^\sigma\rangle$ \\
    2 & $\langle\{V^{\mu\alpha},f_{+}^{\rho\sigma}\}\nabla_\alpha u^\nu\rangle$ \\
    3 & $i\langle\{V^{\mu\nu},f_{+}^{\rho\sigma}\}\chi_-\rangle$ \\
    4 & $i\langle V^{\mu\nu}[f_{-}^{\rho\sigma},\chi_+]\rangle$ \\
    5 &$\langle\{\nabla_\alpha V^{\mu\nu},f_{+}^{\rho\alpha}\} u^\sigma\rangle$\\
    6 & $\langle\{\nabla_\alpha V^{\mu\alpha},f_{+}^{\rho\sigma}\}u^\nu\rangle$\\
    7 & $\langle\{\nabla^\sigma V^{\mu\nu},f_{+}^{\rho\alpha}\}u_\alpha\rangle$ \\ \hline
    \end{tabular}
    \caption{Odd-intrinsic parity operators with a vector resonance $V$, a vector current $J$ and a light pseudoscalar, $P$. 
    The common factor $\varepsilon_{\mu\nu\rho\sigma}$ is omitted in all operators.}
    \label{tab:OVJPops}
\end{table}

\textbf{$\bullet$ Operators with two vector resonance fields:}\\[1ex]

These contributions are of the following type:
\begin{equation}
\mathcal{L}_{VV}^{\mathrm{odd}}=\sum_{j=1}^4 d_j \mathcal{O}^j_{VVP}\,,
\end{equation}
where the only relevant operators in our case are (the common factor $\varepsilon_{\mu\nu\rho\sigma}$ is omitted)
\begin{equation} \label{eq_VVP}
\mathcal{O}_{VVP}^1=\langle\{V^{\mu\nu},V^{\rho\alpha}\}\nabla_\alpha u^{\sigma}\rangle\,,\quad \mathcal{O}_{VVP}^2=i\langle\{V^{\mu\nu},V^{\rho\sigma}\}\chi_-\rangle\,,\quad \mathcal{O}_{VVP}^3=\langle\{\nabla_\alpha V^{\mu\nu},V^{\rho\alpha}\} u^{\sigma}\rangle\,.
\end{equation}
Among these, $\mathcal{O}_{VVP}^2$ breaks unitary flavor symmetry.\\

\textbf{$\bullet$ Operators with excited vector resonance fields:}\\[1ex]

For the second vector multiplet, $V^\prime$, we repeat all operators that we had for the first one, $V$. 
That is, we add eqs.~(\ref{eq_Vmatrix}) to (\ref{eq_VVP})) for the excited vector mesons. In this way, we will have the new parameters (denoted by a prime) $M_{V^\prime},e^{V^\prime}_m,F_{V^\prime},\lambda_{V^\prime},\lbrace c_i^\prime\rbrace_{i=1}^7,\lbrace d_j^\prime\rbrace_{j=1}^3$.

Additionally, we will also have new $VV^\prime P$ interactions, which are~\cite{Mateu:2007tr}\\
\begin{equation}\label{eq_LVV'}
        \mathcal{L}_{VV^\prime}^{\mathrm{odd}}=\sum_{j=a,b,c,d,e,f} d_j \mathcal{O}_{VV^\prime P}^j\,,
\end{equation}
with the $\mathcal{O}_{VV^\prime P}^j$ shown in table~\ref{tab:mixedvvprime}.
\begin{table}[h!]
    \centering
    \begin{tabular}{|c | c |}\hline
    $j$ & $\mathcal{O}^j_{VV'P}$   \\ \hline
     $a$    &  $\langle \left\{V^{\mu\nu},V'^{\rho\alpha} \right\}\nabla_\alpha u^\sigma \rangle$\\
     $b$    &  $ \langle \left\{V^{\mu\alpha},V'^{\rho\sigma} \right\}\nabla_\alpha u^\nu \rangle$\\
     $c$    &  $\langle \left\{\nabla_\alpha V^{\mu\nu},V'^{\rho\alpha} \right\} u^\sigma \rangle$\\ 
     $d$    &  $\langle \left\{\nabla_\alpha V^{\mu\alpha},V'^{\rho\sigma} \right\} u^\nu \rangle$\\ 
     $e$    &  $\langle \left\{\nabla^\sigma V^{\mu\nu},V'^{\rho\alpha} \right\} u_\alpha \rangle$\\ 
     $f$    &  $i\langle \left\{V^{\mu\nu},V'^{\rho\sigma}\right\}\chi_-\rangle$\\
    \hline
    \end{tabular}
    \caption{Odd-intrinsic parity operators with one vector meson resonance, one excited vector meson resonance, and a pseudo-Goldstone boson~\cite{Mateu:2007tr}. 
    A common $\varepsilon_{\mu\nu\rho\sigma}$ factor is omitted in all operators.}
    \label{tab:mixedvvprime}
\end{table}

\textbf{$\bullet$ Operators with pseudoscalar resonance fields:}\\[1ex]

The pseudoscalar resonances, $P^\prime$ (with analogous flavor structure to the $\varphi$ mesons), give subleading contributions via their mixing with the lightest pseudo-Goldstones, $\varphi$, that are suppressed as $m_P^2/m_{P^\prime}^2$. 
However, the $P^\prime$ multiplet is crucial to recover the QCD-ruled short-distance behaviour on the $VVP$ Green's function~\cite{Kampf:2011ty,Roig:2013baa}. 
Because of this, we must also take into consideration the pieces
\begin{eqnarray}\label{eq_LPprime}
\begin{aligned}
\Delta \mathcal{L}_P^{\mathrm{even}}&=\frac{1}{2}\langle \nabla_\mu P' \nabla^\mu P' \rangle+id_m\langle P' \chi_-\rangle\,,\\
\Delta \mathcal{L}_P^{\mathrm{odd}}&=\varepsilon_{\mu\nu\rho\sigma}\langle \kappa_5^P \left\{f^{\mu\nu}_+,f^{\rho\sigma}_+\right\}P'+\kappa_3^{PV}\left\{V^{\mu\nu},f^{\rho\sigma}_+\right\}P'+\kappa^{PVV}V^{\mu\nu}V^{\rho\sigma}P'\rangle\,,
\end{aligned}
\end{eqnarray}
which need to be repeated for $V\to V^\prime$. 
The operator with coefficient $d_m^{(\prime)}$ vanishes in the chiral limit.

There will also be a $VV^\prime P^\prime$ operator~\cite{Kadavy:2020hox}, 
\begin{equation}
        \Delta\mathcal{L}_{VV^\prime P^\prime}^{\mathrm{odd}}=\kappa^{VV^\prime P^\prime} \langle \{V^{\mu\nu},V^{\prime\mu\nu}\}P^\prime\rangle\,,
        \label{VVprimeP}
\end{equation}
that we need to include for consistency.

We will briefly comment next on the flavor breaking induced by our $\mathcal{L}_{\rm{R\chi T}}$, eq.~(\ref{eq_LRChT}), in the resonance sector,\footnote{Chiral symmetry breaking in $\mathcal{L}_{\mathrm{no}\;R}$ happens as in $\chi$PT.} where we account for $\mathcal{O}(m_P^2)$ corrections in interaction vertices, resonance masses and field renormalizations~\cite{Guevara:2018rhj}.

The $e^V_m$ contribution, cf.~eq.~(\ref{eq_Kin}), yields the following pattern of masses for the lightest non-strange neutral vector mesons
\begin{eqnarray}\label{eq_Vmasseswithsymmbreak}
M_\rho^2=M_\omega^2=M_V^2-4e_m^V m_\pi^2\,,\quad M_\phi^2=M_V^2-4e_m^V \Delta_{2K\pi}^2\,,
\end{eqnarray}
with $\Delta_{2K\pi}=2m_K^2-m_\pi^2$. 
Analogous relations are obtained for the $V^\prime$ multiplet.
We will work under the simplifying assumption that their flavor structure is analogous, so that $e_m^{V^\prime} = e_m^V \frac{M_{V^\prime}^2}{M_V^2}$.
The similar shifts for the $P^\prime$ mesons will not be needed, as the induced corrections are subleading, namely $\mathcal{O}(m_P^4)$.
Our cutting of the infinite tower of resonances per set of quantum numbers that are predicted in the large-$N_C$ limit to a few of them implies that the $N_C\to\infty$ masses get shifted as a result. 
Consequently, the masses will be used as free fit parameters and will not be fixed to their physical values.
The uncertainty on $a_\mu^{\rm{HLbL-pole}}$ associated to cutting the infinite spectrum to a few of the lightest states cannot be large, but it will anyway be relevant at the current level of precision. 
The main reason lies in the kernel needed to evaluate this contribution to the $g_\mu-2$, that is strongly dominated by the $[0.1,1]$ GeV$^2$ region (for both photon virtualities), providing $\sim85\%$ of the full results. 
This is considered in the assessment of theory uncertainties of our R$\chi$T approach in section~\ref{sec_Cuttingtower}.

The $V^0$-$\gamma$ transitions will be shifted due to the $\lambda_V$ term in eq.~(\ref{eq.Lagrangin1Reven}). This will imply the following changes
\begin{equation}
(\rho^0/\omega)-\gamma: F_V\to F_V+8m_\pi^2\lambda_V\,,\quad \phi-\gamma: F_V\to F_V+8\Delta_{2K\pi}^2\lambda_V\,,
\end{equation}
with an analogous primed version for the $V^{\prime0}$-$\gamma$ vertices.

\section{Transition form factors in R$\chi$T}\label{sec_TFFs}
\begin{figure}
\centering\includegraphics[width=0.325\textwidth]{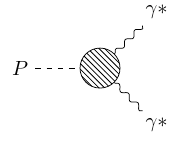}
\caption{Schematic representation of the $P\to\gamma^{*}\gamma^{*}$ ($P=\pi^{0},\eta,\eta^{\prime}$) Transition Form Factor.}
\label{fig:piTFF} 
\end{figure}

\begin{figure}[h!]
\centering
\begin{subfigure}[h]{0.325\textwidth}
    \centering
    \includegraphics[width=0.725\textwidth]{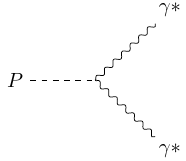}
    \caption{Local}
    \label{figFeynmanDiagsa}
\end{subfigure}   
 \hfill 
\begin{subfigure}[h]{0.325\textwidth}
    \centering
    \includegraphics[width=0.725\textwidth]{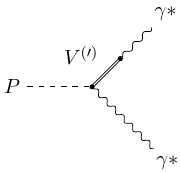}
    \caption{One-resonance exchange}
    \label{figFeynmanDiagsb}
    \end{subfigure}  
 \hfill      
    \begin{subfigure}[h]{0.325\textwidth}
    \centering
    \includegraphics[width=0.725\textwidth]{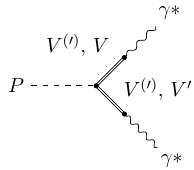}
    \caption{Two-resonances exchange}
    \label{figFeynmanDiagsc}
    \end{subfigure}  
         \caption{Different contributions to the $P\gamma^{*}\gamma^{*}$ vertex.}
        \label{figFeynmanDiags}
\end{figure}

The most general transition amplitude between an on-shell $P=\pi^0/\eta/\eta^{\prime}$ meson and two generally off-shell photons with virtualities (polarizations) $q_1^2(\epsilon_1^*)$ and $q_2^2(\epsilon_2^*)$ is
\begin{equation}
    \mathcal{M}_{P\gamma^*\gamma^*}=ie^2\varepsilon^{\mu\nu\rho\sigma}q_{1\mu}q_{2\nu}\epsilon_{1\rho}^*\epsilon_{2\sigma}^* \mathcal{F}_{P\gamma^*\gamma^*}(q_1^2,q_2^2)\,,
    \label{TFFdefinition}
\end{equation}
which is given in terms of the $P$ transition form factor $\mathcal{F}_{P\gamma^*\gamma^*}(q_1^2,q_2^2)$, fulfilling $\mathcal{F}_{P\gamma^*\gamma^*}(q_1^2,q_2^2)=\mathcal{F}_{P\gamma^*\gamma^*}(q_2^2,q_1^2)$, due to Bose symmetry under the exchange of the identical photons. 
This form factor is illustrated in fig.~\ref{fig:piTFF}.
There are three types of contributions to $\mathcal{F}_{P\gamma^*\gamma^*}(q_1^2,q_2^2)$: local terms, and one- and two-resonance exchange diagrams, as shown in fig.~\ref{figFeynmanDiags}.

The different contributions to $\mathcal{F}_{P\gamma^*\gamma^*}(q_1^2,q_2^2)$ considering only one vector resonance multiplet are described minutely in ref.~\cite{Guevara:2018rhj}, so we will not detail them here. Nonetheless, it is important to quote the redefinition of combinations of the coupling constants appearing in \cite{Guevara:2018rhj} that will also enter our results:~\footnote{This redefinition of coupling constants applies for both vector meson resonance multiplets, so the $c^\prime$s and the $d^{\prime}$s are also redefined in the same way.}
\begin{subequations}
\begin{equation}
c_{1235}\to c_{1235}^*=c_1+c_2+8c_3^*-c_5\,,
\end{equation}
\begin{equation}
d_{123}\to d_{123}^*=d_1+8d_2^*-d_3\,,
\end{equation}
\begin{equation}
d_{abcf}\to d_{abcf}^*=d_a+d_b-d_c+8d_f^*\,.
\end{equation}
\end{subequations}

The contributions from the second vector multiplet can be obtained by simply adding the same terms with primed couplings. We will concentrate here on the new terms that come from the $VV^\prime P$ contributions to the two-resonances exchange diagrams, like in fig.~\ref{figFeynmanDiagsc}. 
After that, we will quote the expressions of the $P$ form factors before requiring short-distance constraints (whence the remaining contributions, can be read off easily).

The contributions coming from $\mathcal{L}_{VV^\prime P}$, eq.~(\ref{eq_LVV'}), read
\begin{eqnarray}\label{eq_V'partpi}
\begin{aligned}
\mathcal{F}^{VV^\prime P}_{\pi^0\gamma^*\gamma^*}(q_1^2,q_2^2)&=
     \frac{2}{3F_\pi}\Bigg\lbrace
(F_V+8m_\pi^2\lambda_V)(F_{V'}+8m_\pi^2\lambda_{V'})\left(m_\pi^2 d_{abcf}
+d_{abc}q_1^2+d_{abcd}q_2^2\right)\\[1ex]
     &\times\left(\frac{1}{(M_\rho^2-q_1^2)(M_{\omega'}^2-q_2^2)}+\frac{1}{(M_{\rho'}^2-q_1^2)(M_{\omega}^2-q_2^2)}\right)+(q_1\leftrightarrow q_2)\Bigg\rbrace\,,
\end{aligned}
\end{eqnarray}
where we defined the combinations of couplings
\begin{equation}
d_{abcf}=d_a+d_b-d_c+8d_f\,,\quad
d_{abc}=d_a-d_b+d_c\,,\quad  
d_{abcd}=-d_a+d_b+d_c-2d_d\,,
\end{equation}
and
\begin{eqnarray}\label{eq_V'parteta}
\mathcal{F}^{VV^\prime P}_{\eta \gamma^*\gamma^*}(q_1^2,q_2^2)&=& \frac{2}{3F
}\Bigg\lbrace \Bigg[5C_q\frac{2}{3}(F_V+8m_\pi^2\lambda_V)(F_{V^\prime}+8m_\pi^2\lambda_{V^\prime})\frac{d_{abcf}
m_\eta^2+d_{abc}q_1^2+d_{abcd}q_2^2-8d_f
\Delta_{\eta\pi}^2}{(M_{\rho\prime}^2-q_1^2)(M_{\rho}^2-q_2^2)}\nonumber\\
&&+(\lbrace5C_q\to-\sqrt{2}C_s,\Delta_{\eta\pi}\to-\Delta_{2K\pi\eta},\rho^{(\prime)}\to\phi^{(\prime)}\rbrace)\Bigg]+(q_1\leftrightarrow q_2)\Bigg\rbrace\,,
\end{eqnarray}
where we introduced 
$\Delta_{\eta\pi}^2=m_\eta^2-m_\pi^2$ and $\Delta_{2K\pi\eta}^2=2m_K^2-m_\pi^2-m_\eta^2$. 
The pion wave-function renormalization has changed the global $1/F$ factor to $1/F_\pi$, for $N_C\to\infty$~\cite{Bernard:1991zc, Sanz-Cillero:2004hed,Guo:2014yva}, in eq.~(\ref{eq_V'partpi}). In this equation. there is no $\phi$ contribution because ideal mixing for the vector meson resonances was used. For the $\eta^{(\prime)}$, $\rho,\omega,\phi$ contribute; However, the isospin symmetry limit leads to $M_{\rho^{(\prime)}}=M_{\omega^{(\prime)}}$ so we have chosen to use $\rho^{(\prime)}$'s mass and propagator only.
For the $\eta^{(\prime)}$ cases, the dependencies are on $C_{q/s}^{(\prime)}/F$, which are functions of ${\lbrace f,\theta\rbrace}_{8/0}$, see eqs.~(\ref{eq_2mixingangles}).

As always, an analogous expression is obtained for $\mathcal{F}_{\eta^\prime \gamma^*\gamma^*}(q_1^2,q_2^2)$ by replacing $C_q\to C_q^\prime$, $C_s\to -C_s^\prime$ and $m_\eta\to m_{\eta^\prime}$, and the $\eta^{(\prime)}$ expressions reduce to the $\pi^0$ one in the $U(3)$ flavor symmetry limit. 
In the chiral limit, our results for the $P$ transition form factors reproduce those in refs.~\cite{Kampf:2011ty,Roig:2014uja}. 
Neglecting the contributions involving the $V^\prime$ multiplet we recover the results in ref.~\cite{Guevara:2018rhj}.

Since we want to make use of the short-distance constraints on the $VVP$ Green's function derived in refs.~\cite{Kampf:2011ty,Roig:2014uja}, the effect of the pseudoscalar resonance multiplet $P^\prime$ needs to be accounted for. 
The operator with coefficient $d_m$ (from $\Delta \mathcal{L}_P^{\mathrm{even}}$ in eq.~(\ref{eq_LPprime})) introduces a mixing between the $P^\prime$ and $\varphi$ states, proportional to quark masses, starting at $\mathcal{O}(m_P^2)$. 
As a result, the contribution from the $P^\prime$ states can be introduced by simply rescaling couplings as follows~\cite{Guevara:2018rhj}:
\begin{subequations}\label{eq_redefinitionsP}
    \begin{equation}
        C_7^W\to C_7^{W*}=C_7^W+\frac{4 d_m \kappa^P_5}{3 M_{P^\prime}}\,,
    \end{equation}
    \begin{equation}
        c_3\to c_3^*=c_3+\frac{d_m M_V\kappa^{PV}_3}{M_{P^\prime}^2}\,,\footnote{A short-distance constraint on $\kappa^{PV}_3$ was found in ref.~\cite{Kampf:2011ty}. We will not use it, like in ref.~\cite{Guevara:2018rhj}, 
        as it does no longer hold when the $V^\prime$ and $P^\prime$ multiplets are also considered.}
    \end{equation}
    \begin{equation}
        d_2\to d_2^*=d_2+\frac{d_m \kappa^{PVV}}{2M_{P^\prime}^2}\,,
    \end{equation}
    \begin{equation}
        d_f\to d_f^*=d_f+\frac{d_m \kappa^{PVV^\prime}}{2M_{P^\prime}^2}\,,
    \end{equation}
\end{subequations}
where the last redefinition was not proposed before, to our knowledge.
The remaining couplings do not need shifting. 

\section{Short distance constraints}\label{sec_SDCs}

We will demand that our form factors, $\mathcal{F}_{P\gamma^*\gamma^*}(q_1^2,q_2^2),\,P=\pi^0,\eta,\eta^\prime$, satisfy the high-energy requirements stemming from QCD~\cite{Nesterenko:1983ef,Novikov:1983jt,Brodsky:1973kr,Lepage:1980fj}:
\begin{subequations}\label{eq_SDCspi0}
\begin{equation}
    \lim_{Q^2\to\infty} -Q^2\mathcal{F}_{\pi^0\gamma^*\gamma^*}(-Q^2,-Q^2)=\frac{2F_\pi}{3}\,,
    \end{equation}
    \begin{equation}\lim_{Q^2\to\infty} -Q^2\mathcal{F}_{\pi^0\gamma^*\gamma^*}(-Q^2,0)=2F_\pi\,,
\end{equation}
\end{subequations}
with $q^2=-Q^2$.
At leading order in the perturbative expansion, the results for the $\eta^{(\prime)}$ mesons are obtained multiplying those for the $\pi^0$, eqs.~(\ref{eq_SDCspi0}), by $\frac{5C_q-\sqrt{2}C_s}{3}\left(\frac{5C_q^\prime+\sqrt{2}C_s^\prime}{3}\right)$, respectively. We comment on $\alpha_S$ and higher-order operator product expansion (OPE) corrections below eq.~(\ref{eq_lastSDC}).

We will apply these constraints first in the chiral limit, and then including $\mathcal{O}(m_P^2)$ corrections. 
Our results will be independent of considering corrections at this order to the LO values for the ${C}_{q/s}^{(\prime)}$ mixing coefficients, so we will keep them unexpanded throughout.

In addition to demanding the leading ultraviolet behaviour in the singly and doubly virtual limits of the transition form factors, we will also require the additional constraints derived from the short-distance analysis of the VVP Green’s function. By matching the leading terms of the QCD OPE for infinite virtualities, within the chiral and large-$N_C$ limits, Kampf and Novotny \cite{Kampf:2011ty} obtain that the following (linear combinations of) constants vanish
\begin{equation}
   C_{22}^W=C_7^{W*}=c_{125}=c_{125}^\prime=c_{1235}=c_{1235}^\prime=0\,,
\end{equation}
where we first used here 
$c_{125}=c_1-c_2+c_5$.
Their remaining results are compatible with our findings that will be detailed next:

\begin{itemize}
    \item Doubly Virtual $\pi^0$-TFF:
    \begin{itemize}
        \item  $\mathcal{O}(Q^0),\,\mathcal{O}(m_P^0)$: 
        \begin{equation}
            c_{1256}^\prime=-M_{V^\prime}\frac{M_V N_C+32 \sqrt{2}\pi^2 F_V c_{1256}}{32\sqrt{2}\pi^2F_{V^\prime}M_V}\,,  
        \end{equation}    
        where $c_{1256}^{(\prime)}=c_1^{(\prime)}-c_2^{(\prime)}-c_5^{(\prime)}+2c_6^{(\prime)}$ appears (primed combinations of constants are defined analogously below).\\
        \item  $\mathcal{O}(Q^0),\,\mathcal{O}(m_P^2)$: 
        \begin{equation}
            \lambda_V=\lambda_{V^\prime}=0\,.  
        \end{equation}
        \item $\mathcal{O}(Q^{-2}),\,\mathcal{O}(m_P^0)$:
        \begin{equation}
            c_{1256}=\frac{8d_3F_V+4d_{abc}F_{V^\prime}-\frac{F_\pi^2 M_V^2}{F_V(M_V^2-M_{V^\prime}^2)}}{4\sqrt{2}M_V}\,.
        \end{equation}
        \item  $\mathcal{O}(Q^{-2}),\,\mathcal{O}(m_P^2)$:
        \begin{equation}
            c_{1235}^*=-e_m^V\frac{8d_3F_V+4d_{abc}F_{V^\prime}-\frac{F_\pi^2 M_V^2}{F_V(M_V^2-M_{V^\prime}^2)}}{\sqrt{2}M_V}\,.
        \end{equation}
    \end{itemize}

    \item Singly Virtual $\pi^0$-TFF:
    \begin{itemize}
        \item $\mathcal{O}(Q^0),\,\mathcal{O}(m_P^0)$: 
        \begin{equation}
            d_3^\prime=\frac{-M_V^2M_{V^\prime}^2N_C -32 \pi^2 d_{abcd}F_V F_{V^\prime}M_V^2-64\pi^2d_3F_V^2M_{V^\prime}^2-32\pi^2d_{abc}F_V F_{V^\prime} M_{V^\prime}^2}{64 \pi^2 F_{V^\prime}^2M_V^2}\,.
        \end{equation}
        \item $\mathcal{O}(Q^0),\,\mathcal{O}(m_P^2)$: 
        \begin{eqnarray}
        \begin{aligned}
            c_{1235}^{*\prime}=&-\frac{M_{V^\prime}}{8\sqrt{2}\pi^2F_{V^\prime}M_V^4}(-e_m^{V^\prime}M_V^4N_C-64\pi^2 d_3 e_m^{V^\prime}F_V^2M_V^2-32\pi^2d_{abc}e_m^{V^\prime}F_VF_{V^\prime}M_V^2\\[1ex]
            &+64\pi^2d_3e_m^V F_V^2 M_{V^\prime}^2+32\pi^2d_{abc}e_m^V F_V F_{V^\prime}M_{V^\prime}^2+8\sqrt{2}\pi^2 c_{1235}^* F_V M_V M_{V^\prime}^2)\,.
        \end{aligned}
        \end{eqnarray} 
        \item $\mathcal{O}(Q^{-2}),\,\mathcal{O}(m_P^0)$:
        \begin{small}
        \begin{equation}
            d_{abc}=-\frac{-M_V^2M_{V^\prime}^4N_C+4\pi^2[4d_{abcd}M_V^2F_V F_{V^\prime}(M_V^2-M_{V^\prime}^2)+16 d_3F_V^2M_{V^\prime}^2(M_V^2-M_{V^\prime}^2)+5F_\pi^2M_V^2M_{V^\prime}^2]}{16\pi^2 F_V F_{V^\prime}M_{V^\prime}^2(M_V^2-M_{V^\prime}^2)}\,.
        \end{equation}
        \end{small}
        \item $\mathcal{O}(Q^{-2}),\,\mathcal{O}(m_P^2)$:
        \begin{equation}
            d_{123}^{\prime*}=-\frac{F_V[2d_{123}^*F_VM_{V^\prime}^2+d_{abcf}^* F_{V^\prime}(M_V^2+M_{V^\prime}^2)]}{2F_{V^\prime}^2M_V^2}\,.
        \end{equation} 
    \end{itemize}
We note that the relation $e_m^{V^\prime}=e_m^V \frac{M_{V^\prime}^2}{M_V^2}$, which stems from our assumption of identical flavor structure for the $V$ and $V^\prime$ nonets, precisely cancels the $\mathcal{O}(Q^0),\,\mathcal{O}(m_P^4)$ terms in the singly virtual asymptotic limit (in fact they are cancelled to all orders in $m_P^2$). We anyway recall that our computation only retains $\mathcal{O}(m_P^2)$ corrections.

We find additional constraints, coming from either the $\eta$ or $\eta^\prime$ form factors, which are:
    \item Doubly Virtual $\eta$-TFF:
    \begin{itemize}
        \item $\mathcal{O}(Q^0)$,$\mathcal{O}(m_P^0)$:
        \begin{equation}
            C_8^W=0\,.
        \end{equation}
        \item $\mathcal{O}(Q^{-2})$,$\mathcal{O}(m_P^2)$:
        \begin{equation}
            c_3^*=e_m^V \frac{16 \pi^2 d_{abcd}  F_V F_{V^\prime}M_V^2(M_V^2-M_{V^\prime}^2)+32\pi^2d_3 F_V^2M_{V^\prime}^2(M_V^2-M_{V^\prime}^2)+24\pi^2F_\pi^2M_V^2M_{V^\prime}^2}{32\sqrt{2}\pi^2 F_V M_V M_{V^\prime}^2 (M_V^2-M_{V^\prime}^2)}\,.
        \end{equation}
    \end{itemize}
    \item Singly Virtual $\eta$-TFF:
    \begin{itemize}
        \item $\mathcal{O}(Q^{0})$,$\mathcal{O}(m_P^2)$:
        \begin{equation}
            c_3^{\prime*}=\frac{M_{V^\prime}^3(\sqrt{2} e_m^V M_V N_C - 128\pi^2 c_3^* F_V) }{128 \pi^2 F_{V^\prime}M_V^3}\,.
        \end{equation}
        \item $\mathcal{O}(Q^{-2})$,$\mathcal{O}(m_P^2)$:
        \begin{equation}\label{eq_lastSDC}
            d_2^{\prime*}=-\frac{F_V[2d_2^* F_V M_{V^\prime}^2+d_f^* F_{V^\prime}(M_V^2+M_{V^\prime}^2)]}{2 F_{V^\prime}^2M_V^2}\,.
        \end{equation}
    \end{itemize}
\end{itemize}

We remark that our setting provides $\pi^0,\eta,\eta^{\prime}$ transition form factors which, in the chiral limit, follow the proportionality $1:\frac{5C_q-\sqrt{2}C_s}{3}:\frac{5C_q^\prime+\sqrt{2}C_s^\prime}{3}$. 
This implies that it is not able to accommodate the corrections to eqs.~(\ref{eq_SDCspi0}) due to the anomalous 
dimension of the singlet axial current~\cite{Leutwyler:1997yr}, which are relevant for the $\eta^{(\prime)}$ cases. 
Our fit results, however, will not hint to any need for improving on this point presently. Higher order corrections to the first of eqs.~(\ref{eq_SDCspi0}) have been computed using the OPE, and multiply it by $\left(1-\frac{8}{9}\frac{\delta_P^2}{Q^2}\right)$. 
For the $\pi^0$, $\delta_\pi^2=0.20(2)$ GeV$^2$, determined from QCD sum rules \cite{Novikov:1983jt}. For the $\eta^{(\prime)}$ the corresponding values have not been computed, although it is reasonable to expect that they deviate from the $\pi^0$ result by typical $U(3)$ (and large-$N_C$) breaking corrections, $\lesssim30\%$~\cite{Masjuan:2017tvw}.

Our short-distance constraints on the R$\chi$T parameters are compatible with those found in $\tau^-\to P^-[\gamma]\nu_\tau$~\cite{Guo:2010dv, Arroyo-Urena:2021nil,Arroyo-Urena:2021dfe} and $\tau^-\to (V P)^-\nu_\tau$ decays~\cite{Guo:2008sh}. 
They are consistent with those derived studying the $\tau^-\to (KK\pi)^-\nu_\tau$~\cite{Dumm:2009kj} and $\tau^-\to\eta^{(\prime)}\pi^-\pi^0\nu_\tau$ \cite{GomezDumm:2012dpx} decays provided $F_V=\sqrt{3}F$~\cite{Roig:2013baa}, a relation that is also favored in $\tau^-\to (\pi\pi\pi)^-\nu_\tau$ decays~\cite{Dumm:2009va, Shekhovtsova:2012ra, Nugent:2013hxa}, which is driven by the axial-vector current.

We will not quote the $\mathcal{F}_{P \gamma^* \gamma^*}$ form factors obtained after applying the short-distance constraints discussed previously, but after employing the next definitions.
First, we introduce the following barred couplings:\footnote{$\Bar{d}_{123}$ and $\Bar{d}_2$ were already used in ref.~\cite{Guevara:2018rhj}.}
\begin{subequations}\label{eq_barredcouplings}
\begin{equation}
        \Bar{d}_{123}=\frac{F_V^2}{3F_\pi^2} d_{123}^*\,,
\end{equation}   
\begin{equation}
    \Bar{d}_{abcf}=\frac{F_V F_{V^\prime}}{6F_\pi^2}d_{abcf}^*\,,
\end{equation}   
\begin{equation}
    \Bar{d}_{2}=\frac{F_V^2}{3F_\pi^2}d_{2}^*\,,
\end{equation}   
\begin{equation}
    \Bar{d}_{f}=\frac{F_V F_{V^\prime}}{6F_\pi^2} d_{f}^*\,,
\end{equation}   
\begin{equation}
    \Bar{d}_{3} =\frac{F_V^2}{3F_\pi^2} d_{3}.
\end{equation}   
\end{subequations}

Further, the dependence of our results on $\Bar{d}_{123}(\Bar{d}_{2})$ and $\Bar{d}_{abcf}(\Bar{d}_{f})$ suggests us to introduce their following convenient combinations
\begin{subequations}\label{eq_ds1ds2}
    \begin{equation}
        d_{s1}=\left(1-\frac{M_{V}^2}{M_{V^\prime}^2}\right)\left(\Bar{d}_{abcf}+\frac{M_{V^\prime}^2}{M_{V}^2}\Bar{d}_{123}\right)\,,
    \end{equation}
    \begin{equation}
        d_{s2}=\left(1-\frac{M_{V}^2}{M_{V^\prime}^2}\right)\left(\Bar{d}_{f}+\frac{M_{V^\prime}^2}{M_{V}^2}\Bar{d}_{2}\right)\,,
    \end{equation}
\end{subequations}
which appear for single and double virtuality, and 
\begin{subequations}
    \begin{equation}
        d_{d1}=\left(1-\frac{M_{V^\prime}^2}{M_{V}^2}\right)\left(\Bar{d}_{abcf}+\Bar{d}_{123}\right)\,,
    \end{equation}
    \begin{equation}
        d_{d2}=\left(1-\frac{M_{V^\prime}^2}{M_{V}^2}\right)\left(\Bar{d}_{f}+\Bar{d}_2\right)\,,
    \end{equation}
   \label{2Vcouplings} 
\end{subequations}
that enter only for double virtuality. It is also advantageous to employ
    \begin{equation}\label{eq_dd3}
        d_{d3}=\left(1-\frac{M_{V^\prime}^2}{M_{V}^2}\right)^2 \Bar{d}_{3},
    \end{equation}
that is only sensitive to the doubly virtual photon case.

Altogether, this enables to recast the $\mathcal{F}_{P \gamma ^* \gamma^*}$ form factors as\begin{equation}
    \begin{aligned}
        \mathcal{F}_{\pi^0 \gamma^*\gamma^*}(q_1^2,q_2^2)=&\Bigg\lbrace96\pi^2F_\pi^2 (m_\pi^2 M_\rho^2 M_{\rho^\prime}^2 d_{s1}+ m_\pi^2 q_1^2 q_2^2 d_{d1}+ 2 M_\rho^2 q_1^2 q_2^2 d_{d3})+N_CM_{V'}^2 M_{\rho\prime}^2(q_1^2q_2^2-M_\rho^4)\\
        &+4F_\pi^2\pi^2q_1^2q_2^2(q_1^2+q_2^2-2M_{\rho}^2)+24F_\pi^2\pi^2 M_{\rho'}^2\left[(q_1^2+q_2^2)M_\rho^2-2q_1^2q_2^2\right] \Bigg\rbrace\\
        &/\left[24 \pi^2 F_\pi(M_\rho^2-q_1^2)(M_\omega^2-q_2^2)(M_{\rho'}^2-q_1^2)(M_{\omega'}^2-q_2^2) \right] + (q_1 \leftrightarrow q_2) \,,  
    \end{aligned}
    \label{piTFFfinal}
\end{equation}
and
\begin{equation}
\begin{aligned}
    &\mathcal{F}_{\eta \gamma ^* \gamma^*}(q_1^2,q_2^2)=\frac{5C_q}{18F\pi^2 (M_\rho^2-q_1^2)(M_\rho^2-q_2^2)(M_{\rho^\prime}^2-q_1^2)(M_{\rho^\prime}^2-q_2^2)}\times\\
    &\Bigg\lbrace \pi^2 F_\pi^2 \Bigg[24  m_\eta^2d_{s1} M_\rho^2M_{\rho^\prime}^2-192 d_{s2} M_\rho^2 M_{\rho^\prime}^2 \Delta_{\eta\pi}^2+24  m_\eta^2 q_1^2 q_2^2 d_{d1}-192 q_1^2 q_2^2 d_{d2}\Delta_{\eta\pi}^2 +48 q_1^2 q_2^2 d_{d3} M_\rho^2 \\
    &- \left( q_1^2 q_2^2(2M_\rho^2-q_1^2-q_2^2)+6M_{\rho\prime}^2\left(2q_1^2q_2^2-M_\rho^2\left(q_1^2+q_2^2\right)\right)\right)\Bigg]+N_C \left(-\frac{M_{V^\prime}^2}{4}\right)\left( M_{\rho^\prime}^2(M_\rho^4-q_1^2q_2^2)\right)\Bigg\rbrace\\
    &+(q_1\leftrightarrow q_2)+(5 C_q \to -\sqrt{2}C_s,\, \rho\leftrightarrow \phi,\, \rho^\prime\leftrightarrow \phi^\prime,\,\Delta_{\eta\pi}^2\to -\Delta_{2K\eta\pi}^2)\,.
\end{aligned}
\label{etaTFFfinal}
\end{equation}
The $\eta^\prime$ form factor is obtained by replacing $C_q\to C_q^\prime$, $C_s\to -C_s^\prime$ and $m_\eta\to m_{\eta^\prime}$ in eq.~(\ref{etaTFFfinal}).\

As in the $\pi^0$ case (see the related explanation below eq.~(\ref{eq_V'parteta})), the $\eta^{(\prime)}$ transition form factors neither depend on the pion decay constant in the chiral limit, $F$. These are functions of $C_q^{(\prime)}/F$ and $C_s^{(\prime)}/F$, which in turn depend just on the two mixing angles $\theta_{8/0}$ and decay constants $f_{8/0}$, (\ref{eq_2mixingangles}).

\section{Form factor data analysis}\label{sec_DataAnalysis}

We start this section recalling the criteria required in the White Paper for taking an evaluation of $a_\mu^{P-\rm{pole,HLbL}}$ into account, before dwelling on the details of our fits to data in section \ref{sec_Fits}. 
These conditions were that (quoting from ref.~\cite{Aoyama:2020ynm}):
\begin{enumerate}
\item in addition to the transition form factor normalization given by the real-photon decay widths, also high-energy constraints must be fulfilled;
\item at least the spacelike experimental data for the singly-virtual TFF must be reproduced;
\item systematic uncertainties must be assessed with a reasonable procedure.
\end{enumerate}

As for the doubly-virtual transition form factor,
the experimental data is still very scarce as there is only one measurement by BaBar consisting of five data points for the $\eta^{\prime}$ transition form factor with relatively large uncertainties. 
Therefore, we require our doubly-virtual transition form factors to be not only in accord with the BaBar data for the $\eta^{\prime}$ but also in line with the Lattice-QCD data released by the Budapest–Marseille–Wuppertal (BMW) collaboration for the $\pi^{0},\eta$ and $\eta^{\prime}$ doubly-virtual TFFs.
As in the singly-virtual case, systematic uncertainties must be also assessed reasonably. We suggest that, for the second White Paper, one should demand --in addition to the three points above-- 
consistency between data-driven $P$ TFF and lattice QCD results, particularly for double virtuality, where these constitute now the best input.

\subsection{Fit to transition form factors data}\label{sec_Fits}

We will start recapitulating the parameters dependence of our $P$ transition form factors, eqs.~(\ref{piTFFfinal}) and (\ref{etaTFFfinal}), with appropriate substitutions for the $\eta^\prime$ case.

In principle, a couple of parameters are needed to specify the masses within each $V$ multiplet, ${M_V,e_m^V}$, and their primed counterparts. However, our assumption of identical flavor structure for both nonets erases the dependence on $e_m^{V^\prime}$. 
Therefore, the corresponding spectra will be specified by three independent parameters, that we choose to be $M_V,e_m^V,M_{V^{\prime}}$. 
In addition to these, we will have the four parameters associated to the $\eta^{(\prime)}$ mixing, our choice being the decay constants and mixing angles ${\lbrace f,\theta\rbrace}_{8/0}$. 
The final five fitted parameters will be those specifying the functional dependence for single and double photon virtuality: $d_{s1,d1,d3}$ for $P=\pi^0$ and also $d_{s2,d2}$ for the $\eta^{(\prime)}$, for a total of $12$ fitted parameters. 
Among them, only $M_{V^{(\prime)}}$ and $d_{d3}$ are $U(3)$-symmetric, while the rest break this flavor symmetry.

Since our focus is on the $P$-pole contributions to $a_\mu^{\rm{HLbL}}$, we will only fit spacelike data ($q^2\leq0$). 
The timelike region is more involved within R$\chi$T, because resonance widths are needed, which is a next-to-leading order effect in the $1/N_C$ expansion. 
Although it is phenomenologically clear that this is the leading next-to-leading order effect, other terms at this order may not be negligible and complicate the treatment. Additionally, radiative corrections can be more important in the timelike region~\cite{Kampf:2018wau}.\\

We will use the following data:
\begin{itemize}
    \item The decay widths $\Gamma(P\to\gamma\gamma)$, corresponding to the transition form factors evaluated for null virtuality (real photons),
    \begin{equation}
        \Gamma(P \to \gamma \gamma)=\frac{(4\pi \alpha)^2}{64\pi}m_P^3 |\mathcal{F}_{P\gamma\gamma}(0,0)|^2\,,
    \label{eq_radwidth}
    \end{equation}
are helpful in the characterization of the $\eta$-$\eta^\prime$ mixing and constitute the normalization which receives chiral corrections for low virtualities. 
These inputs are taken from the PDG~\cite{Workman:2022ynf}.\footnote{We are not taking $\Gamma(\eta^\prime\to\gamma\gamma)$ as a separate data point, since it uses LEP data that we are fitting, see next item.}
    \item Transition form factor data from the BaBar~\cite{BaBar:2009rrj, BaBar:2011nrp}, Belle~\cite{Belle:2012wwz}, CELLO~\cite{CELLO:1990klc},  CLEO~\cite{CLEO:1997fho} and LEP~\cite{L3:1997ocz} experiments, for the singly virtual case.

\item There is only one measurement for double virtuality, by BaBar, in the $\eta^{\prime}$ channel~\cite{BaBar:2018zpn}, consisting of five points (which are insufficient to reliably fit our three parameters $d_{d1,d2,d3}$). 
We will increase the sensitivity to the doubly virtual regime by supplementing BaBar's with Lattice QCD results~\cite{Gerardin:2023naa, ExtendedTwistedMass:2022ofm, ExtendedTwistedMass:2023hin}. 
Specifically, from the $z$-expansion performed by the BMW collaboration in~\cite{Gerardin:2023naa}, we generate three points (at $Q_1^2=Q_2^2=0.1\, ,1.0$, and $4.0$ GeV$^2$) for all three $P$ mesons.\footnote{Given that the $z$-expansion has 6 parameters (with their corresponding correlation) one can generate at most 6 points for having an invertible covariance matrix. 
Four our analysis, we only generate 3 {\it{lattice}} points for each $P$ to avoid high correlations between (neighboring) points~\cite{Workman:2022ynf}. 
We note that 3 is the minimum number of points we need to fix the three doubly virtual parameters. 
Had we used 4, 5 or 6 data points we would have obtained highly correlated data, which implies a (close to) non-invertible covariance matrix, that over-represents the lattice information.}
These points are shown in Table~\ref{tab:LatticePoints}.
We would like to note here that at least the Lattice data for the $\eta$ doubly-virtual transition form factor shall be included in the fits to obtain a satisfactory simultaneous description of all three mesons.
\begin{table}
    \centering
    \begin{tabular}{|l|lll|}
\hline
        $Q_{1}^{2}=Q_{2}^{2}$ [GeV$^{2}$] 
        & 0.1 &  1& 4\\
\hline 
$\pi^{0}$& 0.0194(3) & 0.0475(4) & 0.0514(12)\\
0.1         &1  & 0.2758 & 0.1556\\
1      & 0.2758 &  1& 0.1222\\
4     & 0.1556 & 0.1222 & 1\\
\hline
$\eta$& 0.0158(11) & 0.0440(26) & 0.0474(31)\\
0.1         &1  & 0.6743 & 0.3006\\
1      & 0.6743 &  1& 0.4615\\
4     & 0.3006 & 0.4615 & 1\\
\hline
$\eta^{\prime}$& 0.0251(30) & 0.0920(100) & 0.0934(114)\\
0.1         &1  & 0.8658 & 0.3840\\
1      & 0.8658 &  1& 0.4423\\
4     & 0.3840 & 0.4423 & 1\\
\hline
    \end{tabular}
    \caption{Central values, uncertainties and correlation matrix for the doubly virtual transition for factors, generated at three representative values of $q_{1}^{2}=q_{2}^{2}$ from the BMW results~\cite{Gerardin:2023naa} and used in our fits.}
    \label{tab:LatticePoints}
\end{table}
\end{itemize}

As in ref.~\cite{Guevara:2018rhj}, we will take advantage of stabilization points for the fit, using the results from a previous determination of the ${\lbrace f,\theta\rbrace}_{8/0}$ parameters~\cite{Feldmann:1998vh, Kaiser:2000gs, Schechter:1992iz, Bramon:1997va, Feldmann:1999uf, Escribano:2005qq}\footnote{The correlation between the fitted parameters describing the singly and doubly virtual behavior (now $d_{s1,s2,d1,d2}$) and ${\lbrace f,\theta\rbrace}_{8/0}$ is large, like with a single vector resonance multiplet (as in ref.~\cite{Guevara:2018rhj}), which calls for adding these extra fit points in order to keep the fit within the physical region.}
\begin{eqnarray}\label{eq_ValuesEtaEta'Mixing}
\begin{aligned}
 \theta_8 =(-21.2 \pm 1.6)^\circ\,,\quad& \theta_0 =(-9.2 \pm 1.7)^\circ,\\[1ex] 
 f_8 =(1.26 \pm 0.04)F_\pi = (116.2 \pm 3.7) \mathrm{MeV} &,  
f_0 =(1.17 \pm 0.03)F_\pi = (107.9 \pm 2.8) \mathrm{MeV}\,,
\end{aligned}
\end{eqnarray}
which increase the $\chi^2$ by barely more than unit.

The fits can compensate for the neglected higher $V$ excitations by shifting any of the mass parameters $M_V$ or $M_V^\prime$. We choose to keep $M_\rho$ close to its PDG value by adding it as a point in the cost function.

We will add to the $\chi^2$ a final fit point, corresponding to $\delta_\pi^2=0.20(2)$ GeV$^2$ (see the discussion below eq.~(\ref{eq_lastSDC})).

The cost function for this fit will then be:
\begin{equation}
    \chi^2_{\mathrm{Global}}=\chi^2_{\pi^0_{\mathrm{SV}}}+\chi^2_{\eta^{}_{\mathrm{SV}}}+\chi^2_{\eta^\prime_{\mathrm{SV}}}+\chi^2_{\pi^0_{\mathrm{DV}}}+\chi^2_{\eta^{}_{\mathrm{DV}}}+\chi^2_{\eta^\prime_{\mathrm{DV}}}+\sum_{P}^{\mathrm{Extra Points}}\left(\frac{P_\mathrm{exp}-P_{\mathrm{model}}}{\Delta P_{\mathrm{exp}}}\right)^2,
    \label{chiglobalDV}
\end{equation}
which includes the transition form factor data for single and double virtuality, SV and DV (where the latter incorporate lattice input) for the $P$ channels, and the extra points given by: the $\Gamma(P\to\gamma\gamma)$ decay widths~\cite{Workman:2022ynf}, the decay constants and mixing angles of the $\eta-\eta^\prime$ system (\ref{eq_ValuesEtaEta'Mixing}), $M_\rho$~\cite{Workman:2022ynf}, 
and $\delta_\pi^2=0.20(2)$ GeV$^2$~\cite{Novikov:1983jt}.

Regarding the Lattice data, the correlations between the generated data were also considered in each of the $\chi^2_{P^{\mathrm{LQCD}}_{DV}}$, so their contribution to the reduced $\chi^2$ in eq.~(\ref{chiglobalDV}) is given by:
\begin{equation}
    \chi^2_{P^\mathrm{LQCD}_{DV}} = \sum_{i,j=1}^{3}\left(P^{\mathrm{LQCD}}_i-P^{\mathrm{R\chi T}}_i\right)\left({\rm{Cov}}^{{\rm{LQCD}}}_{ij}\right)^{-1}\left(P^{\mathrm{LQCD}}_j-P^{\mathrm{R\chi T}}_j\right),
\end{equation}
where $P^{{\rm{LQCD}}}_{i(j)}$  and $\left({\rm{Cov}}^{{\rm{LQCD}}}_{ij}\right)^{-1}$ are, respectively, the central value and the inverse of the covariance matrix of the lattice data given in Table~\ref{tab:LatticePoints}.~\footnote{Taking the covariance matrix into account introduces extra degrees of freedom due to having non-diagonal entries different from zero equal. The additional $\mathrm{d.o.f.}$ equal the non redundant entries of the covariance matrix.}

Following previous analyses of the $\mathcal{F}_{P \gamma^*\gamma^*}$ form factors and their contributions to $a_\mu^{\rm{HLbL}}$ \cite{Hoferichter:2020lap,Roig:2014uja, Guevara:2018rhj, Mikhailov:2009kf, Roberts:2010rn, Agaev:2010aq, Brodsky:2011yv, Bakulev:2011rp, Brodsky:2011xx, Stefanis:2012yw, Bakulev:2012nh, Agaev:2012tm, Agaev:2014wna, Raya:2015gva, Eichmann:2017wil,Stefanis:2020rnd} and owing to our preliminary fits for this work, we will not include BaBar $\pi^0$ data in our reference fit, as it is inconsistent with Belle's, that appears more compatible with the predicted (Brodsky-Lepage) asymptotic limit.\footnote{We have checked that the fits are worse if these BaBar data is kept, like in ref.~\cite{Guevara:2018rhj}.}
Additionally, the chiral symmetry relations among the different $P$ form factors data points at the largest measured energies are best satisfied without the BaBar $\pi^0$ data. 
The uncertainty associated to the reliability of the latter is assessed in section~\ref{sec_Cuttingtower}. 

Our preliminary fits show a big correlation (close to one) of the pairs of parameters $\lbrace d_{s1},d_{s2}\rbrace$ and $\lbrace d_{d1},d_{d2}\rbrace$. This motivates us to consider instead their combinations
\begin{subequations}
    \begin{equation}
        d_{s1}=\langle d_{s1}\rangle +\frac{\sigma_{d_{s1}}}{\sqrt{2}}\left(\sqrt{1+r_s}\,r_{s1}-\sqrt{1-r_s}\,r_{s2}\right)\,,
    \end{equation}
    \begin{equation}
        d_{s2}=\langle d_{s2}\rangle +\frac{\sigma_{d_{s2}}}{\sqrt{2}}\left(\sqrt{1+r_s}\,r_{s1}+\sqrt{1-r_s}\,r_{s2}\right)\,,
    \end{equation}
   \label{eq_decorrelation} 
\end{subequations}
and analogously for the doubly virtual case (${}s\to{}d$), in order to minimize their correlations (assuming small interdependences with the rest of the parameters, which is a good approximation).
Eqs.~(\ref{eq_decorrelation}) use the mean values of the preliminary fit ($\langle d_{s1}\rangle, \langle d_{s2}\rangle$) and their uncertainties ($\sigma_{d_{s1}}, \sigma_{d_{s2}}$). 
The new parameters are $r_{s1},r_{s2}$, defined through the original correlation $r_s\sim 1$ (same with ${}s\to{}d$).

\begin{figure}
\vspace{-2em}
\centering\includegraphics[width=0.75\textwidth]{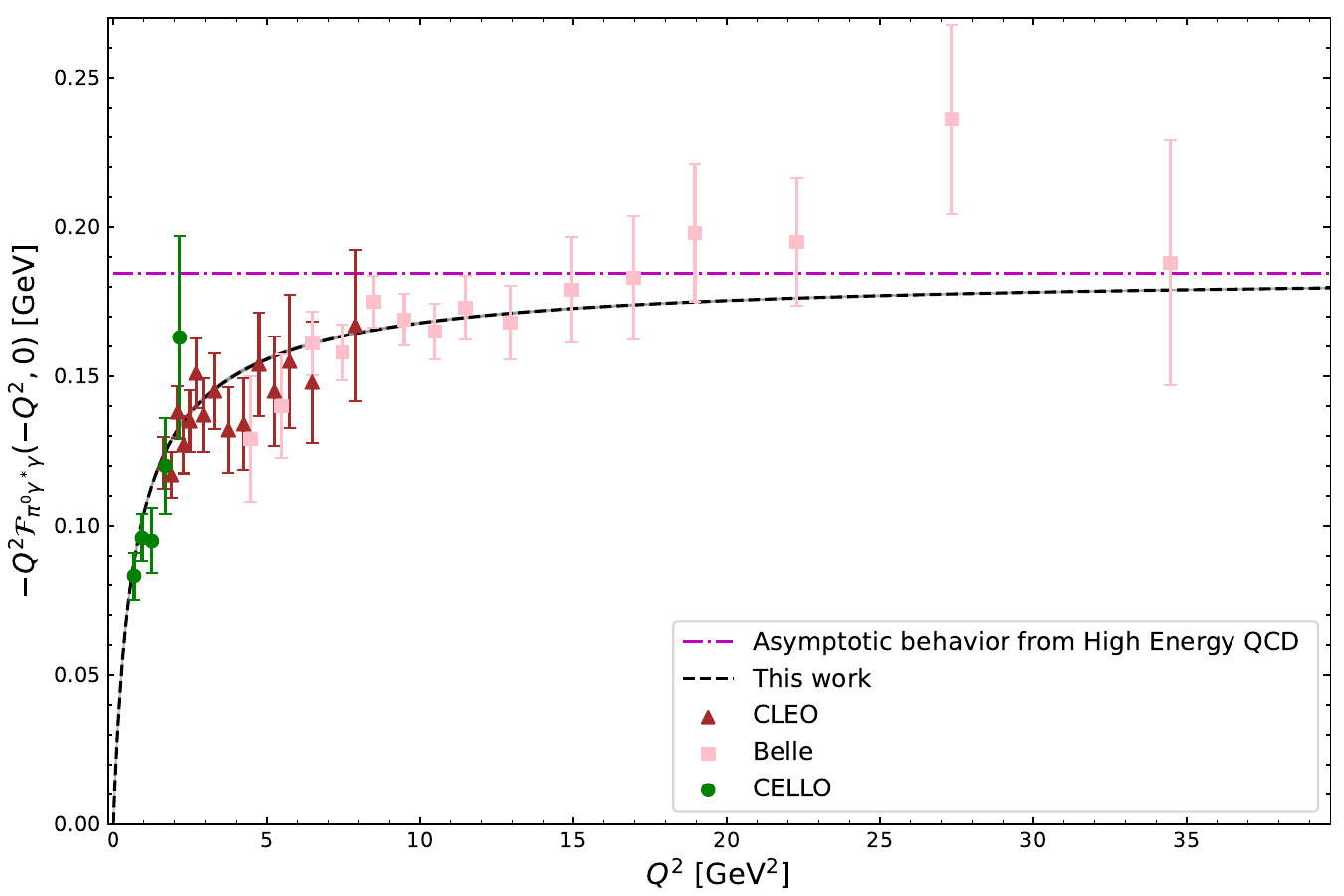}\\[1ex]
\centering\includegraphics[width=0.75\textwidth]{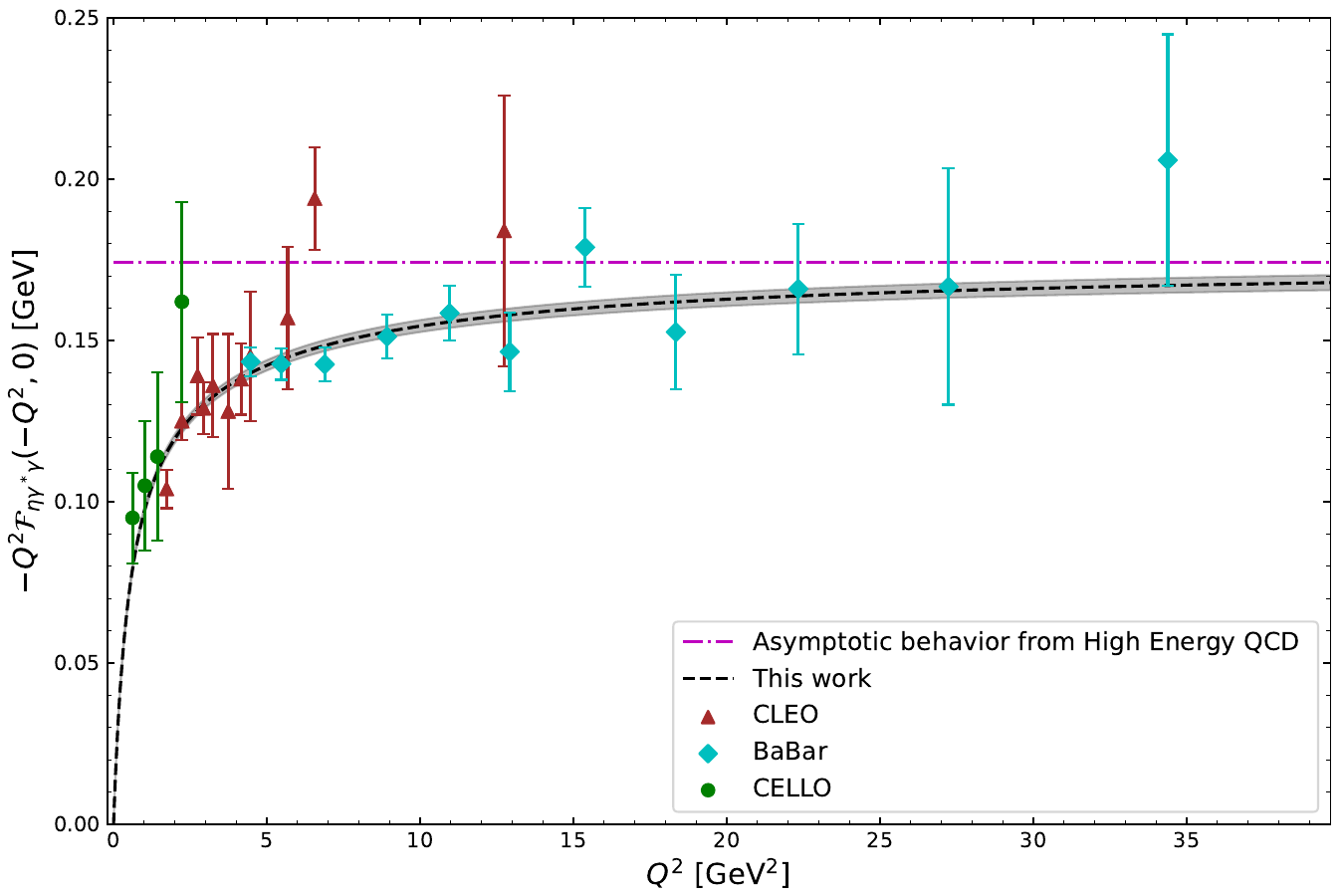}\\[1ex]
\centering\includegraphics[width=0.75\textwidth]{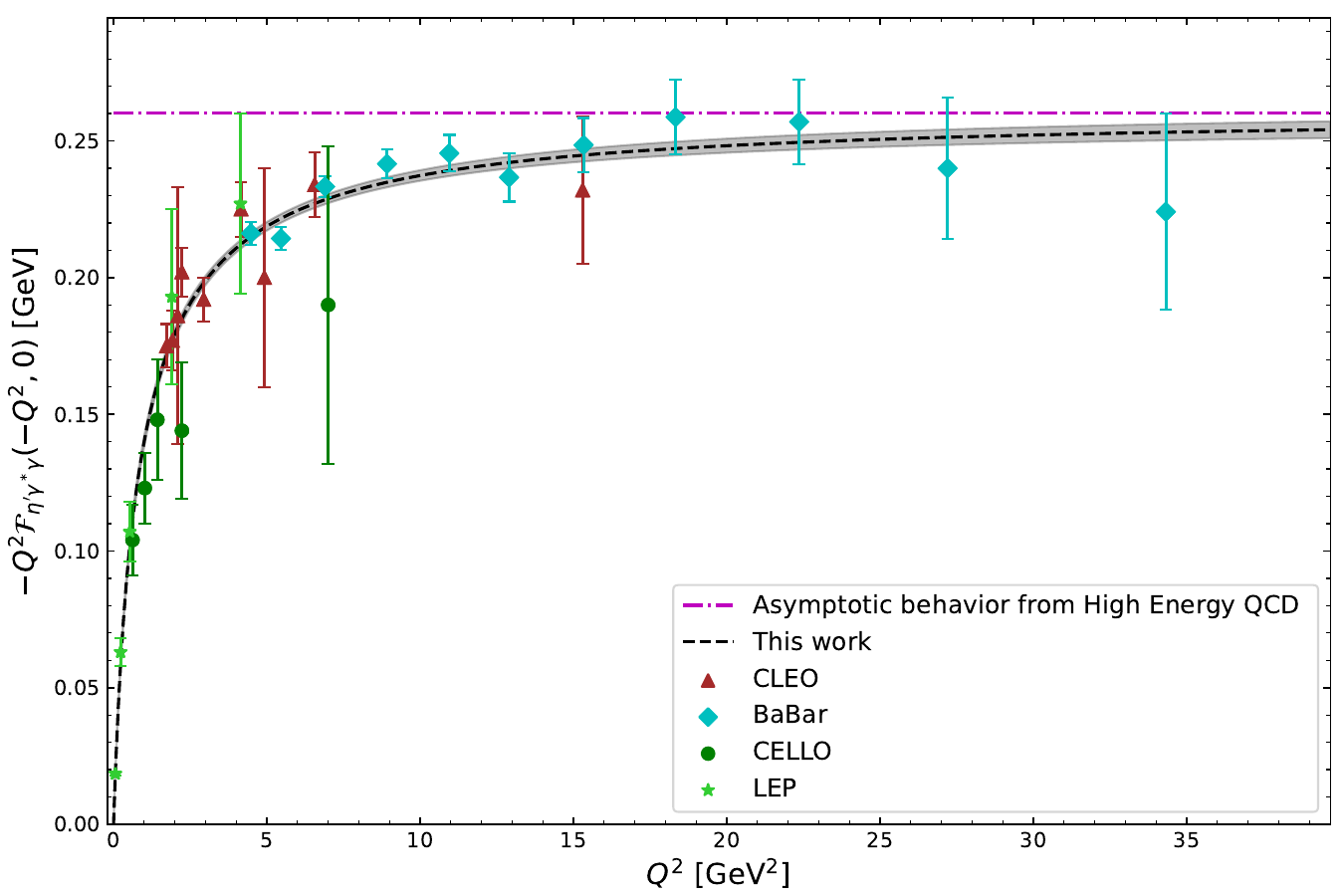}
\caption{The transition form factors corresponding to our best fit results (table~\ref{tab:completefitresults}) are compared to data in the singly virtual regime for $\pi^0$, $\eta$ and $\eta^\prime$. 
We do not show BaBar data for the $\pi^0$ case~\cite{BaBar:2009rrj}, which was not used in our reference fits.}
\label{fig:TFF1V}
\end{figure}

\begin{figure}
\vspace{-1em}
\centering\includegraphics[width=0.75\textwidth]{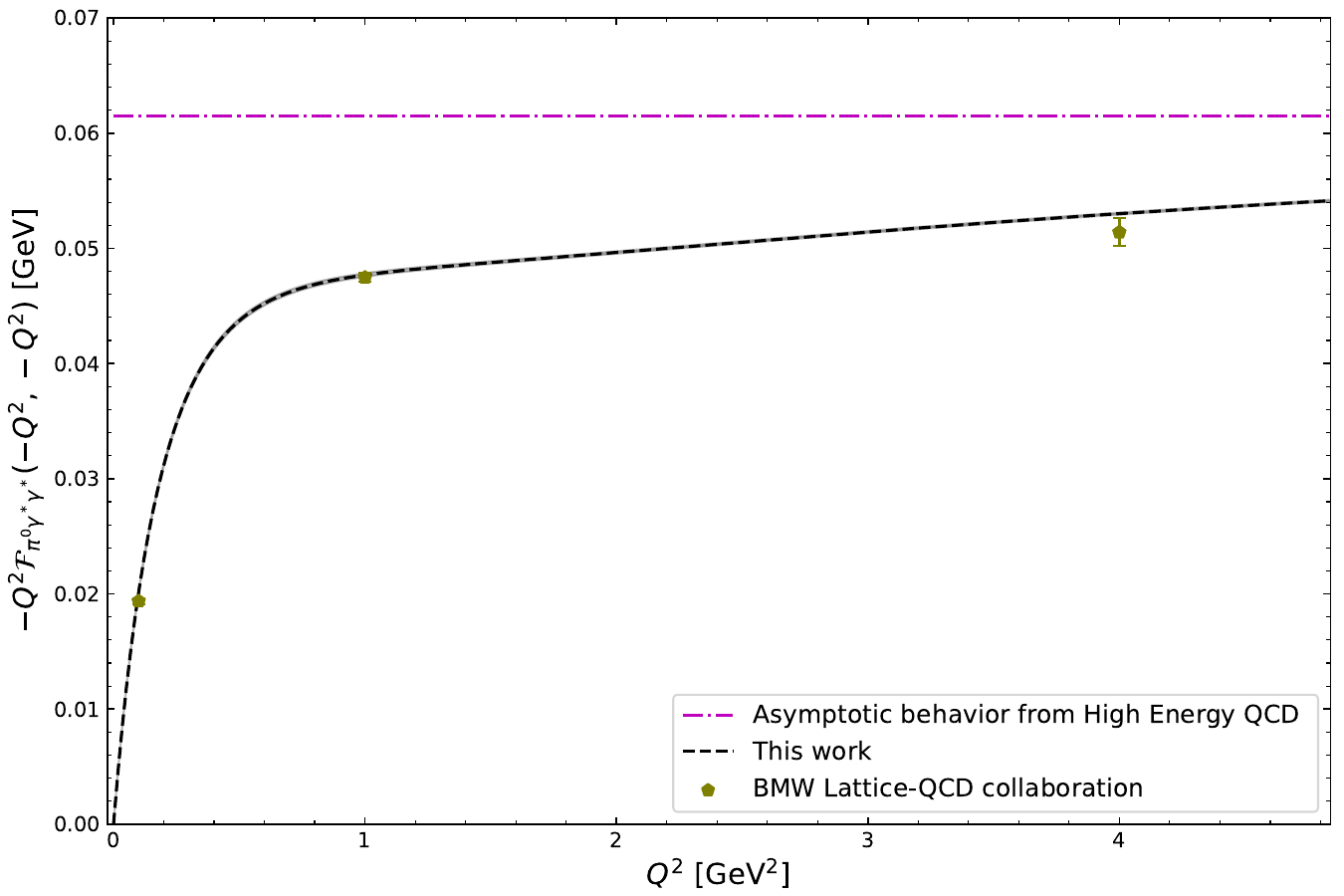}\\[1ex]
\centering\includegraphics[width=0.75\textwidth]{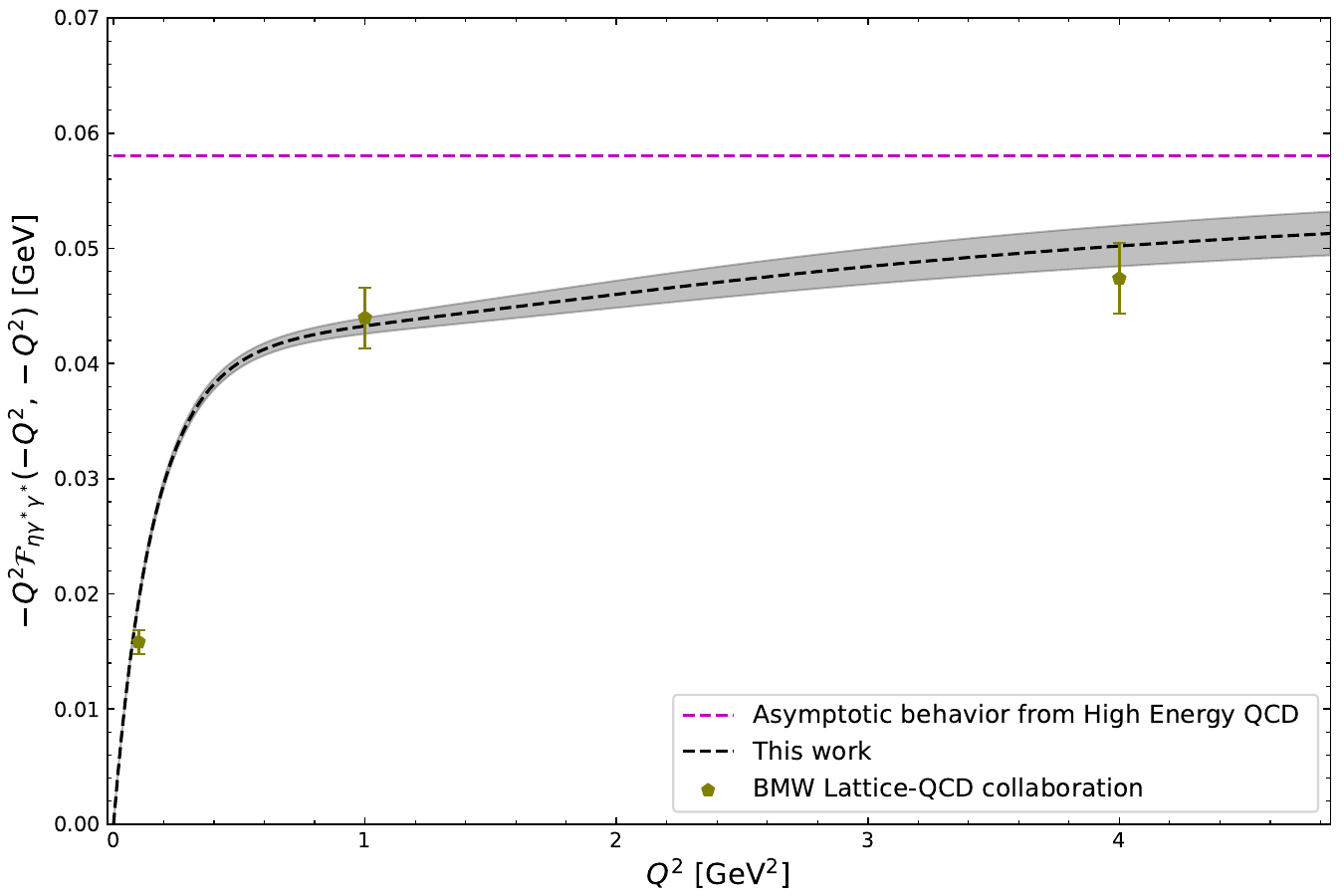}\\[1ex]
\centering\includegraphics[width=0.75\textwidth]{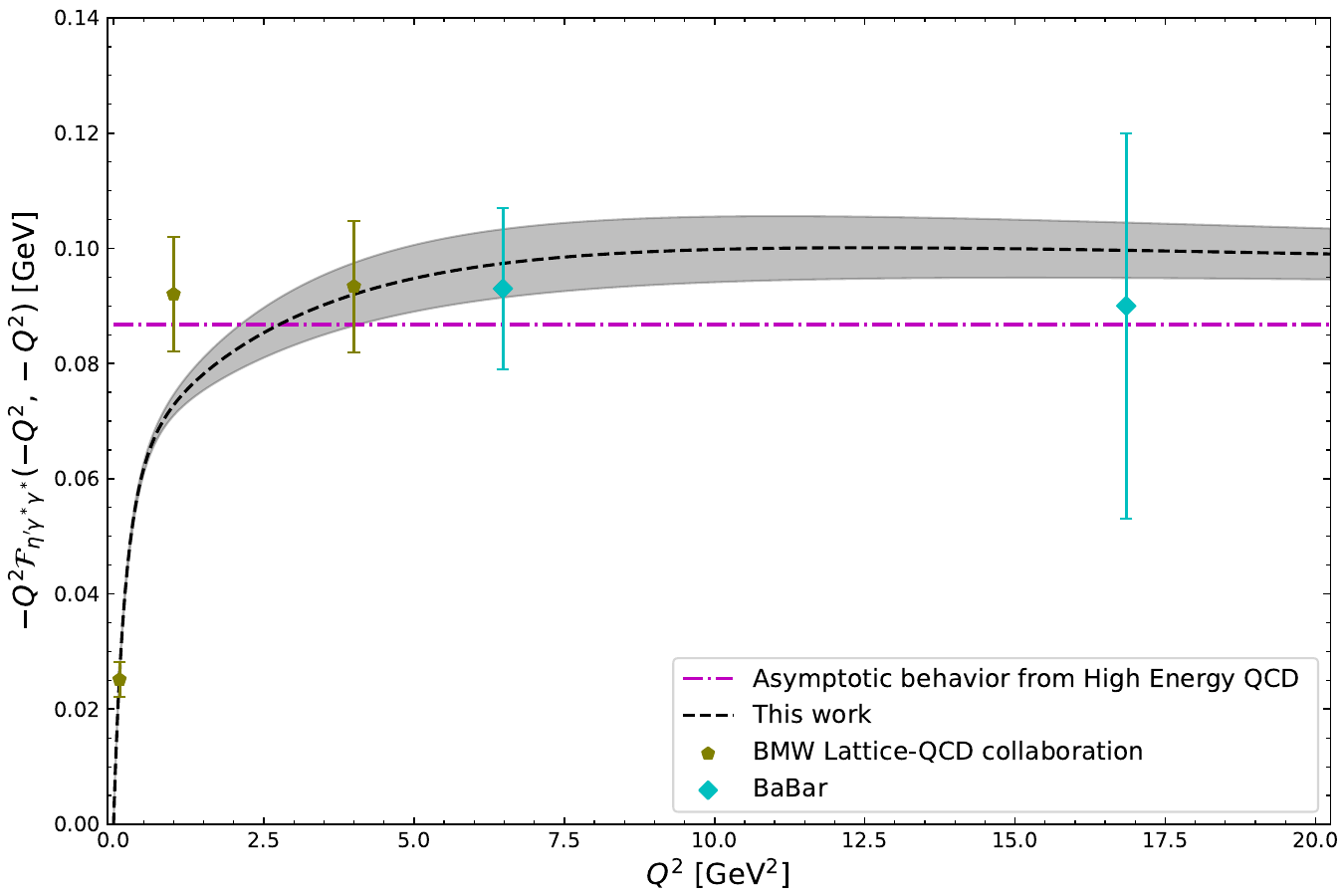}
\caption{The transition form factors corresponding to our best fit results (table~\ref{tab:completefitresults}) are compared to data in the doubly virtual regime for $\pi^0$, $\eta$ and $\eta^\prime$.}
    \label{fig:TFF2V}
\end{figure}

Our best fit results corresponding to the minimization of the cost function (\ref{chiglobalDV}) are given in table \ref{tab:completefitresults} with the correlation matrix shown in table~\ref{tab:fitcorrelations}. 
The comparison of our $\mathcal{F}_{P\gamma^*\gamma^*}(q_1^2,q_2^2)$ to data for single and double virtuality are displayed in figures~\ref{fig:TFF1V} and~\ref{fig:TFF2V}.

\begin{table}[h!]
    \centering
    \begin{tabular}{|l | l|}
    \hline
        Parameter & Fit Result\\
        \hline
        $M_V$ [GeV] &  0.752(2)\\
        $e_m^V$ &  -0.32(4)\\
        $M_{V^\prime}$ [GeV] &  1.933(4)\\
        $r_{s1}$    & 0.0(0.6)\\
        $r_{s2}$    & 0.0(0.9)\\
        $r_{s}$ & 0.9976\\
        $\langle d_{s1}\rangle$ & -1.6(6)\\
        $\langle d_{s2}\rangle$ &  -0.21(7)\\
        $\theta_8$ [$^\circ$]&  -18.5(6) \\
        $\theta_0$  [$^\circ$]& -6.9(1.6)\\
        $f_8$ [MeV]& 118.8(4)\\
        $f_0$ [MeV]& 99.4(1.7)\\
        $r_{d1}$ & 0.0(0.5)\\
        $r_{d2}$ & 0.0(0.9)\\
        $r_d$ & 0.9783\\
        $\langle d_{d1}\rangle$ &-2.8(2.0)\\
        $\langle d_{d2}\rangle$ &-0.31(24)\\
        $d_{d3}$ &-3.48(3)\\
        $\chi^2_{Global}/\mathrm{d.o.f.}$ &148.0/110\\
        	\hline
    \end{tabular}
    \caption{Our best fit results, with uncertainties in parentheses. The input values of the rotations (\ref{eq_decorrelation}) are also given, where the errors in the $\langle d\rangle$ are the $\sigma_d$'s.}
    \label{tab:completefitresults}
\end{table}

\begin{table}[h!]
\begin{tiny}
\centering
\begin{tabular}{|c|c c c c c c c c c c c c|} \hline
& $M_V$ & $e_m^V$ & $M_{V^\prime}$ & $r_{s1}$ & $r_{s2}$ & $\theta_8$ & $\theta_0$& $f_8$& $f_0$ & $r_{d1}$&$r_{d2}$&$d_{d3}$\\ \hline     
$M_V$& 1&$ 0.739 $&$ -0.035 $&$ 0.501 $&$ -0.065 $&$ 0.035 $&$ -0.008 $&$ -0.077 $&$ -0.025 $&$ 0.357 $&$ -0.152 $&$ 0.371 $\\
$e_m^V$&$ 0.739 $& 1&$ -0.106 $&$ 0.614 $&$ -0.120 $&$ -0.009 $&$ 0.019 $&$ -0.099 $&$ -0.053 $&$ 0.300 $&$ -0.145 $&$ 0.344 $\\
$M_{V^\prime}$&$ -0.035 $&$ -0.106 $& 1&$ -0.035 $&$ -0.114 $&$ -0.032 $&$ 0.202 $&$ 0.063 $&$ -0.030 $&$ -0.446 $&$ 0.399 $&$ -0.837 $\\
$r_{s1}$&$ 0.501 $&$ 0.614 $&$ -0.035 $& 1&$ 0.226 $&$ 0.335 $&$ 0.029 $&$ -0.014 $&$ -0.113 $&$ 0.217 $&$ -0.097 $&$ 0.186 $\\
$r_{s2}$&$ -0.065 $&$ -0.120 $&$ -0.114 $&$ 0.226 $& 1&$ 0.185 $&$ -0.219 $&$ 0.089 $&$ 0.431 $&$ 0.026 $&$ 0.211 $&$ 0.082 $\\
$\theta_8$ &$ 0.035 $&$ -0.009 $&$ -0.032 $&$ 0.335 $&$ 0.185 $& 1&$ 0.002 $&$ -0.657 $&$ -0.069 $&$ 0.075 $&$ 0.071 $&$ 0.015 $\\
$\theta_0$&$ -0.008 $&$ 0.019 $&$ 0.202 $&$ 0.029 $&$ -0.219 $&$ 0.002 $& 1&$ -0.001 $&$ 0.605 $&$ -0.069 $&$ -0.041 $&$ -0.187 $\\
$f_8$&$ -0.077 $&$ -0.099 $&$ 0.063 $&$ -0.014 $&$ 0.089 $&$ -0.657 $&$ -0.001 $& 1&$ 0.032 $&$ -0.047 $&$ 0.002 $&$ -0.084 $\\
$f_0$&$ -0.025 $&$ -0.053 $&$ -0.030 $&$ -0.113 $&$ 0.431 $&$ -0.069 $&$ 0.605 $&$ 0.032 $& 1&$ -0.011 $&$ 0.165 $&$ 0.030 $\\
$r_{d1}$&$ 0.357 $&$ 0.300 $&$ -0.446 $&$ 0.217 $&$ 0.026 $&$ 0.075 $&$ -0.069 $&$ -0.047 $&$ -0.011 $& 1&$ -0.152 $&$ 0.140 $\\
$r_{d2}$&$ -0.152 $&$ -0.145 $&$ 0.399 $&$ -0.097 $&$ 0.211 $&$ 0.071 $&$ -0.041 $&$ 0.002 $&$ 0.165 $&$ -0.152 $& 1&$ -0.392 $\\
$d_{d3}$&$ 0.371 $&$ 0.344 $&$ -0.837 $&$ 0.186 $&$ 0.082 $&$ 0.015 $&$ -0.187 $&$ -0.084 $&$ 0.030 $&$ 0.140 $&$ -0.392 $& 1\\
\hline
\end{tabular}
\end{tiny}
\caption{Correlation matrix between the 12 fitted parameters of the best fit.}
\label{tab:fitcorrelations}
\end{table}

We will comment now on the big correlations that can still be seen in table~\ref{tab:fitcorrelations}. 
The redefinition in eq.~(\ref{eq_dd3}) shows that we cannot make any further rotation of parameters to avoid the anticorrelation of $-0.837$ between $M_V^\prime$ and $d_{d3}$.\footnote{Besides, the correlation between them is non-linear, so we could not proceed analogously.} 
We rotated the parameters $d_{s1}$ and $d_{s2}$, according to eq.~(\ref{eq_redefinitionsP}), to minimize their correlation. 
We note that the rotated parameter $r_{s1}$ has big correlations with both $M_V$ and $e^V_m$ (which is not the case for its partner rotated parameter, $r_{s2}$). 
Correlations between the $\eta$-$\eta^\prime$ mixing parameters are slightly larger than in ref.~\cite{Guevara:2018rhj}, but still reasonable. 
The large correlation between $M_V$ and $e^V_m$ is entirely a result of fixing $M_\rho$ to its PDG value. 
In our preliminary fits, floating both parameters independently, their correlation was only $\sim0.22$.\\

We will discuss next the central values of our reference fit:
\begin{itemize}
\item $M_V$ is typically smaller than the results obtained in single resonance approximations ($\sim800$ MeV \cite{Lepage:1980fj,Brodsky:1981rp}) but the value of $M_{\rho(\phi)}$ is still compatible, at 1(2) $\sigma$, with the PDG\cite{Workman:2022ynf}.

\item As commented above, $e^V_m$ is highly correlated with $M_V$ because we are requiring that $M_V^2-4e^V_mm_\pi^2\sim M_\rho^2$, cf. eq.~(\ref{eq_Vmasseswithsymmbreak}). Nevertheless, the value of this flavor symmetry breaking parameter is compatible at less than one standard deviation with the best fit result in ref.~\cite{Guevara:2018rhj}.
\item One should not expect $M_{V^\prime}^2-4e^{V^\prime}_mm_\pi^2\sim M_{\rho^\prime}^2$, because $M_{V^\prime}^2$ (we recall that we are assuming $e_m^{V^\prime}M_V^2=e_m^V M_{V^\prime}^2$) absorbs the effect of the neglected higher $V$ excitations, so its fitted value $M_{V^\prime}\sim 1.933$ GeV appears reasonable.
\item Concerning the $\eta$-$\eta^\prime$ mixing parameters, agreement of our fitted values with the input (\ref{eq_ValuesEtaEta'Mixing}) is acceptable, with differences being $1.2(0.70)\sigma$ for $\theta_{8(0)}$, and $0.4(1.9)\sigma$ for $f_{8(0)}$. 
\item For our best fit values, eq.~(\ref{eq_dd3}) implies that $d_{d3}\sim 30 \Bar{d}_3$, with $\Bar{d}_3\sim d_3$. Short-distance QCD constraints on the $VVP$ Green's function \cite{Kampf:2011ty,Roig:2013baa} determine $d_3\sim-0.126$, corresponding to a $d_{d3}$ with a deviation of less than 10\% from our best fit value.
\item Best fit values for the parameters $r_{s1,s2}$ and $r_{d1,d2}$ are compatible with zero. Little information is known on the couplings of the $V^\prime$ multiplet that enter the definition of the original couplings $d_{s1,s2}$ and $d_{d1,d2}$. If we set them to zero for a rough estimate and take into account that $\Bar{d}_{123,2}\sim {d}_{123,2}$, we can use that $d_{123}\sim1/24$~\cite{Kampf:2011ty,Roig:2013baa} (again from the short-distance behaviour of the $VVP$ Green's function), to estimate that $d_{s1}\sim1/4$ and $d_{d1}\sim5/4$, which agree within 1 and 3$\sigma$  with our best fit. According to refs.~\cite{Chen:2012vw, Dai:2013joa, Chen:2013nna, Chen:2014yta} $d_2=0.08\pm 0.08$, that again yields estimates $d_{s2}$ and $d_{d2}$ in agreement with our best fit results.
\end{itemize}

The comparison of the transition form factors corresponding to our best fit results (table \ref{tab:completefitresults}) to data in the singly virtual regime for the $\pi^0$, $\eta$ and $\eta^\prime$ mesons in fig.~\ref{fig:TFF1V} is very satisfactory but for a few points that seem to be outliers, given the general trend shown by the data. 
At the largest measured virtualities our transition form factors seem to approach faster the Brodsky-Lepage limit than the data, that have large uncertainties there, anyway.

\begin{figure}
\centering\includegraphics[width=0.75\textwidth]{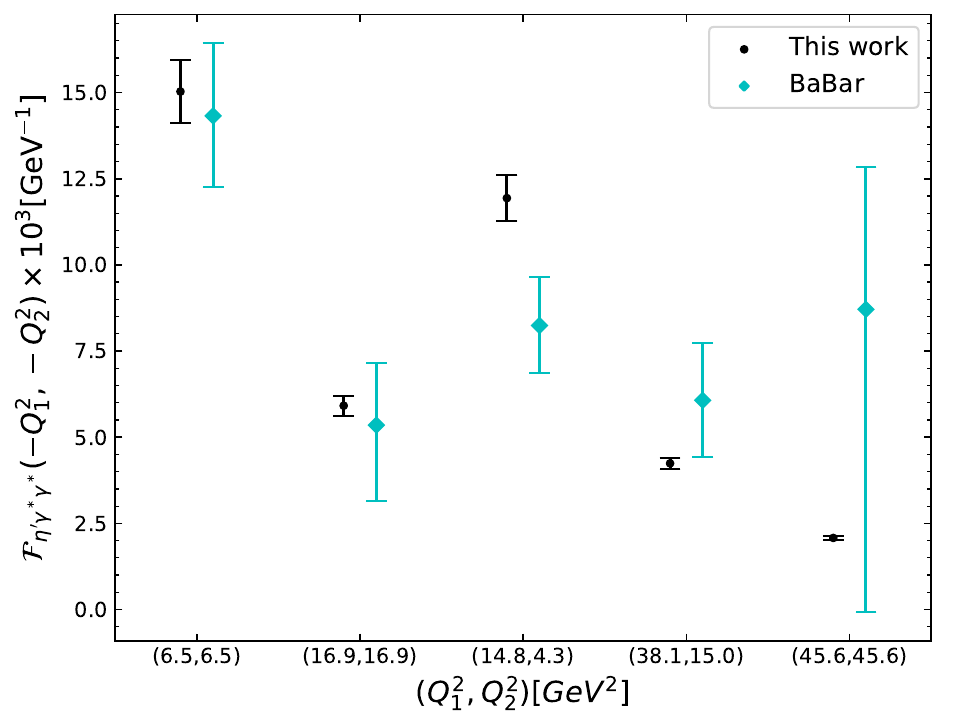}
\caption{BaBar data (cyan diamonds) for the $\eta^{\prime}$ doubly-virtual transition form factor~\cite{BaBar:2018zpn} as compared to our best fit results (black circles) from table~\ref{tab:completefitresults}.}
    \label{fig:FitvsBaBar}
\end{figure}

An analogous comparison for double virtuality is shown in fig.~\ref{fig:TFF2V}, where the relevance of the lattice data is evident -these data are extremely helpful in our analysis to constrain the doubly-virtual parameters $d_{d1,d2,d3}$.
We note that a couple of low-energy points in the $\eta\,(\eta^\prime)$ case are below (above) our curves, although the statistical significance of this tension is moderate. 
As a sanity check, we have verified that the comparison of our best fit results to lattice QCD exhibits a similar pattern in the singly virtual case.\footnote{We cannot add these data to our fit since they would be double counted (they are obtained from the double virtuality data, setting one photon on-shell). 
We could have decided to use this information for the singly virtual case, but it was much more useful to employ it for double virtuality in our case, as we did.} 
In addition, in fig.~\ref{fig:FitvsBaBar} we show a comparison of the standalone BaBar measurements of the $\eta^{\prime}$ doubly-virtual transition form factor, which includes non-diagonal $Q^{2}$ data points, together with our best fit results. 
As seen, the results are in good agreement, while the experimental statistics is not sufficiently precise yet as to further constrain the parameters of double virtuality.

We have extracted the slope parameters, ${\lbrace b,c,d\rbrace}_P$, defined from
\begin{equation}
    \lim_{Q^2\to 0}\mathcal{F}_{P\gamma^*\gamma}(-Q^2,0)=\mathcal{F}_{P\gamma\gamma}(0,0)\left(1-\frac{b_P}{m_P^2}Q^2+\frac{c_P}{m_P^4}Q^4-\frac{d_P}{m_P^6}Q^6+\cdots\right)\,,
\end{equation}
corresponding to our best fit results, and compare them to ref.~\cite{Masjuan:2017tvw} in table~\ref{tab:slopeparametersresults}. 
Like this reference, we do not quote our results for $d_\pi$, because the sensitivity of the data and our fit to this parameter is very small, because of chiral suppression. 
The agreement for $b_P$ (and for $c_\pi$) is very good while the accord with the other parameters shown is quite satisfactory. 
In the last row for each channel, we recall the coefficient of the $\mathcal{O}(1/Q^2)$ term in the single asymptotic (Brodsky-Lepage) limit ($P_\infty$), see eq.~(\ref{SDC-SV}), for which both analyses agree within one standard deviation.

\begin{table}[]
    \centering
    \begin{tabular}{ | l | l | l | c|} \hline 
        Parameter & Our result & Values in \cite{Masjuan:2017tvw} & Difference between both\\ \hline
        $b_\pi$  & 0.03163(16) & 0.0321(19)& $0.2\sigma$\\
        $c_\pi$  & 0.000862(6) &0.00104(22) & $0.8\sigma$\\ 
        $\pi_\infty$
        & $2F_\pi$ & $2F_\pi$ & $0\sigma$\\ \hline
        $b_\eta$  & 0.600(10) & 0.572(8) & $1.6\sigma$\\
        $c_\eta$  & 0.316(7) & 0.333(9) &$1.1\sigma$ \\ 
        $d_\eta$  & 0.164(4) & 0.195(20) & $1.3\sigma$\\         
        $\eta_\infty$[GeV] & 0.174(3) &0.180(12) & $0.4\sigma$\\  \hline  
        $b_{\eta^\prime}$ & 1.26(2) & 1.31(3) & $1.0\sigma$\\
        $c_{\eta^\prime}$ & 1.86(4) & 1.74(9)& $0.9\sigma$\\ 
        $d_{\eta^\prime}$ & 2.82(6) & 2.30(22)& $1.9\sigma$\\         
        $\eta^\prime_\infty$[GeV]& 0.260(4)&0.255(4)& $0.6\sigma$\\ \hline      
    \end{tabular}
    \caption{Low-energy slope parameters ${\lbrace b,c,d\rbrace}_P$ and Brodsky-Lepage parameter ($P_\infty$) from our best fit result and their comparison with the values used in~\cite{Masjuan:2017tvw}.}
    \label{tab:slopeparametersresults}
\end{table}

Finally, it is important to remark that, besides having reproduced the Short-Distance Constraints (SDC) for all 3 particles and having a good agreement with other descriptions at intermediate energies, the decay widths $\Gamma(P\to \gamma \gamma)$ are also in great agreement with the experimental values -we have obtained a value of $7.67(9)$ eV for $\pi^{0}$ (which is a 0.6$\sigma$ difference with respect to~\cite{PrimEx:2010fvg,PrimEx-II:2020jwd}), $497(13)$ eV for the $\eta$ (0.6$\sigma$ difference with~\cite{KLOE-2:2012lws}) and $4.3(2)$ KeV for $\eta^\prime$ (0.2$\sigma$ difference with~\cite{L3:1997ocz}). 
This agreement is noteworthy, given the slight tension (in the $\pi^0$ and $\eta$ cases) between the lattice QCD results for these widths and the PDG values (see also our footnote 13). 
\section{Equivalence with the Canterbury Approximants}\label{sec_CA}
Since the White Paper values~\cite{Aoyama:2020ynm} for the $\eta^{(\prime)}$-pole contributions to $a_\mu^{\rm{HLbL}}$ come from the Canterbury Approximants (CA) analysis of ref.~\cite{Masjuan:2017tvw}, we will translate our R$\chi$T description including two multiplets of vector resonances to the CA language and comment on the corresponding implications in this section.\footnote{Obviously, it is impossible to relate analytically our framework to the dispersive approach~\cite{Hoferichter:2018kwz}, which provides the reference result for the $\pi^0$ contribution~\cite{Aoyama:2020ynm}. 
A dispersive analysis for the $\eta^{(\prime)}$ contributions is in progress~\cite{EtaEtaPDispersive}.}

CA were successfully used for the description of the transition form factors of the non-strange neutral pseudoscalar mesons $\pi^0$, $\eta$ and $\eta^\prime$~\cite{Masjuan:2017tvw} (see ref.~\cite{Escribano:2015vjz} for the extrapolation to the timelike region). 
They are described by a rational function $f(x,y)$ of polynomials which are analytic, symmetric under $x\leftrightarrow y$ and satisfy the accuracy-through-order conditions~\cite{Masjuan:2017tvw,CAs1,CAs2,CAs3,Masjuan:2015lca,Masjuan:2015cjl}. A $C_{M}^N$ is defined then as:
\begin{equation}
    C_M^N(x,y)=\frac{R_N(x,y)}{Q_M(x,y)}=\frac{\sum_{i,j=0}^N a_{i,j}x^iy^j}{\sum_{i,j=0}^M b_{i,j}x^iy^j}\,.
    \label{CAdef}
\end{equation}

Ref.~\cite{Masjuan:2017tvw} mentioned that both $C_{N+1}^N(Q_1^2,Q_2^2)$ and $C_{N}^N(Q_1^2,Q_2^2)$ 
work for describing the high-energy behavior prescribed by perturbative QCD~\cite{Brodsky:1973kr,Lepage:1980fj,Nesterenko:1983ef,Novikov:1983jt}, eqs.~(\ref{eq_SDCspi0}). 
Choosing one or the other depends on whether dropping the last term(s) of the polynomial from $R_N$ or $Q_M$ in eq. (\ref{CAdef}). 
Given this constriction, the convergence and Bose symmetry are guaranteed at arbitrary virtualities for both photons. 
Increasing $N$ and/or $M$ implies incrementing the freedom of the model, so that an optimal choice of them must be done with both freedom and over-fitting in mind. In~\cite{Masjuan:2017tvw}, a $C_2^1$ was used\footnote{The $F_{P\gamma\gamma}(0,0)$ was factored out to provide with a physical meaning the normalization constant, given $a_0=b_0=1$. Besides, the $\beta_{2,2}$ term was dropped -as commented before- in order to correctly account for the high-energy behavior.} (we do not write a subscript $P$ in the $\alpha$ and $\beta$ coefficients in eq.~(\ref{CA12}), although they are different for $\pi^0,\eta,\eta^\prime$):
\begin{equation}
C_2^1(Q_1^2,Q_2^2)=\frac{\mathcal{F}_{P\gamma\gamma}(0,0)\left(1+\alpha_1(Q_1^2+Q_2^2)+\alpha_{1,1}Q_1^2Q_2^2\right)}{1+\beta_1(Q_1^2+Q_2^2)+\beta_{1,1}Q_1^2Q_2^2+\beta_2(Q_1^4+Q_2^4)+\beta_{1,2}Q_1^2 Q_2^2(Q_1^2+Q_2^2)}.
\label{CA12}
\end{equation}
The low-energy behavior, given by the $Q_1^2,Q_2^2\to 0$ expansion~\cite{Escribano:2015yup}, is:
\begin{eqnarray}
\begin{aligned}
    \mathcal{F}_{P\gamma^*\gamma^*}(Q_1^2,Q_2^2)=\mathcal{F}_{P\gamma\gamma}(0,0)&\bigg(1-\frac{b_P}{m_P^2}(Q_1^2+Q_2^2)+\frac{c_P}{m_P^4}(Q_1^4+Q_2^4)\\
&+\frac{a_{P;1,1}}{m_P^4}(Q_1^2Q_2^2)-\frac{d_P}{m_P^6}(Q_1^6+Q_2^6)+\cdots\bigg)\,,
    \label{slope-parameters}
\end{aligned}
\end{eqnarray}
and has been widely studied for all 3 particles (see Table VI from~\cite{Masjuan:2017tvw}). 
The values of $\mathcal{F}_{P\gamma\gamma}(0,0)$, $b_P$ and $c_P$, together with the high-energy behavior constraints
\begin{subequations}
\begin{equation}
    \lim_{Q^2\to \infty}\mathcal{F}_{P\gamma^*\gamma^*}(Q^2,Q^2)=\frac{P_\infty}{3}\left(\frac{1}{Q^2}-\frac{8}{9}\frac{\delta^2_P}{Q^4}\right)+\mathcal{O}(Q^{-6})\,,
    \label{SDC-DV}
\end{equation}    
\begin{equation}
    \lim_{Q^2\to \infty}\mathcal{F}_{P\gamma^*\gamma}(Q^2,0)=\frac{P_\infty}{Q^2}\,,
    \label{SDC-SV}
\end{equation}
\label{SDC}
\end{subequations}
impose 6 restrictions to the form factors in eq.~($\ref{CA12}$), leaving only $\alpha_{1,1}$ as a free parameter.\footnote{For the $\eta^{(\prime)}$ cases, the asymptotic constraint on $P_\infty$ in eq.~(\ref{SDC-SV}) was traded by the low-energy one on $d_P$ (we recall there is no sensitivity to it for $P=\pi^0$), see table~\ref{tab:slopeparametersresults}.} 
However, $\alpha_{1,1}$ could not be fitted therein, since it is sensitive only to the doubly virtual case, for which no data was available by then.\footnote{In ref.~\cite{Masjuan:2017tvw} the range for $\alpha_{P\,1,1}$ was taken so as to  avoid poles for the $C_2^1(Q_1^2,Q_2^2)$ in the spacelike region.} 
Now, there are both experimental data~\cite{BaBar:2018zpn} (only for $\eta^\prime$) and lattice QCD evaluations~\cite{Gerardin:2023naa, ExtendedTwistedMass:2022ofm, ExtendedTwistedMass:2023hin} which can be -as we illustrated in this work- used to generate data in order to complete this description. 
An updated version of the CA study is included in this work, both for its intrinsic interest and for comparing to our results in the previous section.

Our form factors satisfying short-distance QCD constraints, eqs.~(\ref{piTFFfinal}) and (\ref{etaTFFfinal}) -with trivial changes for $\eta\to\eta^\prime$- correspond to a $C_2^2$ and a $C_4^4$ CA, respectively. 
For the $\pi^0$ case the CA coefficients matching our parametrization eq.~(\ref{piTFFfinal}) are given in table~\ref{tab:piCATHScorrespondence}.

\begin{table}[h!]
    \centering
    \begin{tabular}{| l | l |}\hline
    CA Coefficient & R$\chi$T result \\ \hline
    $F_{\pi^0\gamma\gamma}(0,0)$& $\frac{8 d_{s1}F_\pi m_\pi^2}{M_\rho^2 M_{\rho^\prime}^2}-\frac{M_V^2N_C}{12F_\pi M_{\rho}^2 \pi^2}$\\
    $\alpha^\pi_1 F_{\pi^0\gamma\gamma}(0,0)$& $-\frac{2F_\pi}{M_\rho^2M_{\rho^\prime}^2}$\\
    $\alpha^\pi_{1,1} F_{\pi^0\gamma\gamma}(0,0)$& $-\frac{4F_\pi}{M_\rho^4 M_{\rho^\prime}^2}+\frac{2 F_\pi \left(-M_\rho^2+12d_{d1}m_\pi^2+24 d_{d3}M_\rho^2\right)}{3M_\rho^4 M_{\rho^\prime}^4}+\frac{M_V^2 N_C}{12 F_\pi \pi^2 M_\rho^6}$\\
    $\alpha^\pi_2 F_{\pi^0\gamma\gamma}(0,0)$& 0\\
    $\alpha^\pi_{1,2}F_{\pi^0\gamma\gamma}(0,0)$& $-\frac{F_\pi}{3 M_\rho^4 M_{\rho^\prime}^4}$\\
    $\alpha^\pi_{2,2} F_{\pi^0\gamma\gamma}(0,0)$&0\\
    $\beta^\pi_{1}$& $M_\rho^{-2}\left(1+\frac{M_{V}^2}{M_{V^\prime}^2}\right)$\\
    $\beta^\pi_{1,1}$& $M_\rho^{-4}\left(1+\frac{M_{V}^2}{M_{V^\prime}^2}\right)^2$\\
    $\beta^\pi_{2}$& $M_\rho^{-2}M_{\rho^\prime}^{-2}$\\
    $\beta^\pi_{1,2}$& $M_\rho^{-4}M_{\rho^\prime}^{-2}\left(1+\frac{M_{V}^2}{M_{V^\prime}^2}\right)$\\
    $\beta^\pi_{2,2}$& $M_\rho^{-4}M_{\rho^\prime}^{-4}$\\
 \hline
    \end{tabular}
    \caption{Translation of our R$\chi$T result, eq.~(\ref{piTFFfinal}) to CA, for the $\pi^0$.}
    \label{tab:piCATHScorrespondence}
\end{table}

We make the following important remarks concerning the results in table~\ref{tab:piCATHScorrespondence}:
\begin{itemize}
    \item The $C_2^2$ model has 4 more parameters than the $C_2^1$ model used in~\cite{Masjuan:2017tvw}. However, two of them are zero, consequently the degrees of freedom increase by two.
    \item By construction, our model based on R$\chi$T reproduces the dominant behavior of the transition form factors, but the $\delta_\pi^2$ is only included as a fit point in the $\chi^2$, as opposed to the $C_2^1$ case, where it is used to fix a coefficient.
\item The slope parameters from~\cite{Escribano:2015yup} are imposed on the $C_2^1$ up to $c_\pi$. 
In our case (R$\chi$T) they are not imposed, but rather we check (satisfactorily) compatibility with the results for them in ref.~\cite{Masjuan:2017tvw} of our best fit results (see table~\ref{tab:slopeparametersresults}). 
\item For our R$\chi$T results, in the $\pi^0$ case, there are only two independent terms in the denominator of the CA, that we choose as $\beta_{1}$ and $\beta_2$. For the $\eta^{\prime}$ we have $\beta_{1,2,3,4}$, which are independent. 
    The others can be written as $\beta_{i,j}=\beta_i \beta_j$, so they are not quoted in the following.
\item In general, there is only 1 free parameter in $C_2^1$ (3 for $C_2^2$) per channel, which can now be fitted to 3 points for the $\pi^0$ and $\eta$ (from lattice), and 8 for the $\eta^\prime$ (from lattice and the BaBar measurement). 
 In our R$\chi$T description there are 12 parameters which were simultaneously fitted to 122 data points\footnote{We included eight additional stabilization points, see the paragraphs before eq.~(\ref{chiglobalDV}).} in the three channels, from different experiments~\cite{CLEO:1997fho, CELLO:1990klc, L3:1997ocz,BaBar:2009rrj, BaBar:2011nrp,Belle:2012wwz,BaBar:2018zpn} (including PDG~\cite{Workman:2022ynf}) and lattice data, generated from ref.~\cite{Gerardin:2023naa}.
\end{itemize}

For the $\eta^{(\prime)}$ cases, working with such a high order CA ($C_4^4$) is not feasible. 
For illustrative purposes, we obtained the corresponding CA in the chiral limit, which is a $C_2^2$. In fact, the same one describes the three cases ($\pi^0,\,\eta,\,\eta^\prime$), provided we factor out the overall $1:(5C_q-\sqrt{2}C_s)/3:(5C_q'+\sqrt{2}C_s')/3$ dependence ($c_P$). 
The translation of the chiral limit of our results, eqs.~(\ref{piTFFfinal}) and (\ref{etaTFFfinal}), to CA is given in table~\ref{tab:ChiralLimitTHStoCA}.

\begin{table}[h!]
    \centering
    \begin{tabular}{| l | l |}\hline
    CA Coefficient & Chiral THS coefficient \\ \hline
    $F_{P\gamma\gamma}(0,0)$& $-\frac{c_P N_C}{12 F_\pi \pi^2}$\\
    $\alpha_1 F_{P\gamma\gamma}(0,0)$ & $-\frac{2 c_P  F_\pi}{M_V^2 M_{V^\prime}^2}$\\
    $\alpha_{1,1}F_{P\gamma\gamma}(0,0)$ & $c_P\left(\frac{16 \Bar{d}_{3}F_\pi(M_V^2-M_{V^\prime}^2)^2}{M_V^6 M_{V^\prime}^4}+\frac{-\frac{8F_\pi^2(M_V^2+6M_{V^\prime}^2)}{M_{V^\prime}^4}+\frac{N_C}{\pi^2}}{12F_\pi M_V^4}\right)$\\
    $\alpha_{2}F_{P\gamma\gamma}(0,0)$ & 0\\
    $\alpha_{1,2}F_{P\gamma\gamma}(0,0)$ & $-\frac{c_P F_\pi}{3M_V^4 M_{V^\prime}^4}$\\
    $\beta_{1}$& $\left(\frac{1}{M_V^2}+\frac{1}{M_{V^\prime}^2}\right)$\\
    $\beta_{2}$& $\frac{1}{M_V^2 M_{V^\prime}^2}$\\
     \hline
    \end{tabular}
    \caption{Translation of the chiral limit of our R$\chi$T result, eqs.~(\ref{piTFFfinal}) and (\ref{etaTFFfinal}), to CA, for $P=\pi^0,\eta,\eta^\prime$.}
    \label{tab:ChiralLimitTHStoCA}
\end{table}

In our updates of the CA results of ref.~\cite{Masjuan:2017tvw}, to be presented below, we have decided to use the same methodology that was employed in this reference for the $\pi^0$ also for the $\eta^{(\prime)}$. 
That is, we will not trade the constraint on $P_\infty$ (Brodsky-Lepage) by the one on $d_P$ for $P=\eta^{(\prime)}$.\footnote{We decided to do this because it is consistent with the R$\chi$T procedure, see the comparison in table \ref{tab:slopeparametersresults}. 
We have checked that proceeding in complete analogy to ref.~\cite{Masjuan:2017tvw} yields slightly larger $\chi^2$ in the best fit results.} We will also consider the fits using $C_2^2$ enforcing the constraints that ensure that no poles be generated in the spacelike region.

\begin{figure}
\centering\includegraphics[width=0.73\textwidth]{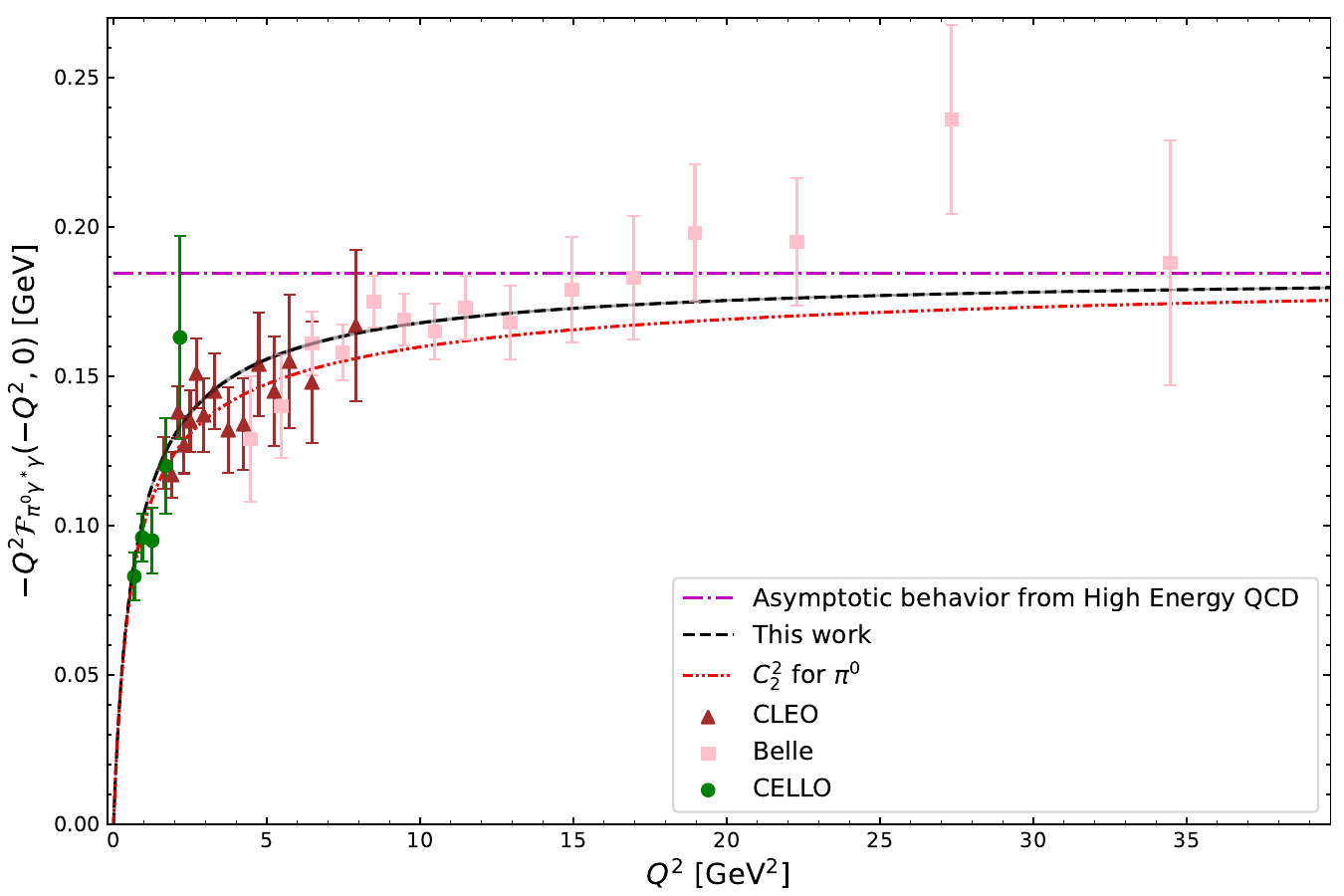}\\[1ex]
\centering\includegraphics[width=0.73\textwidth]{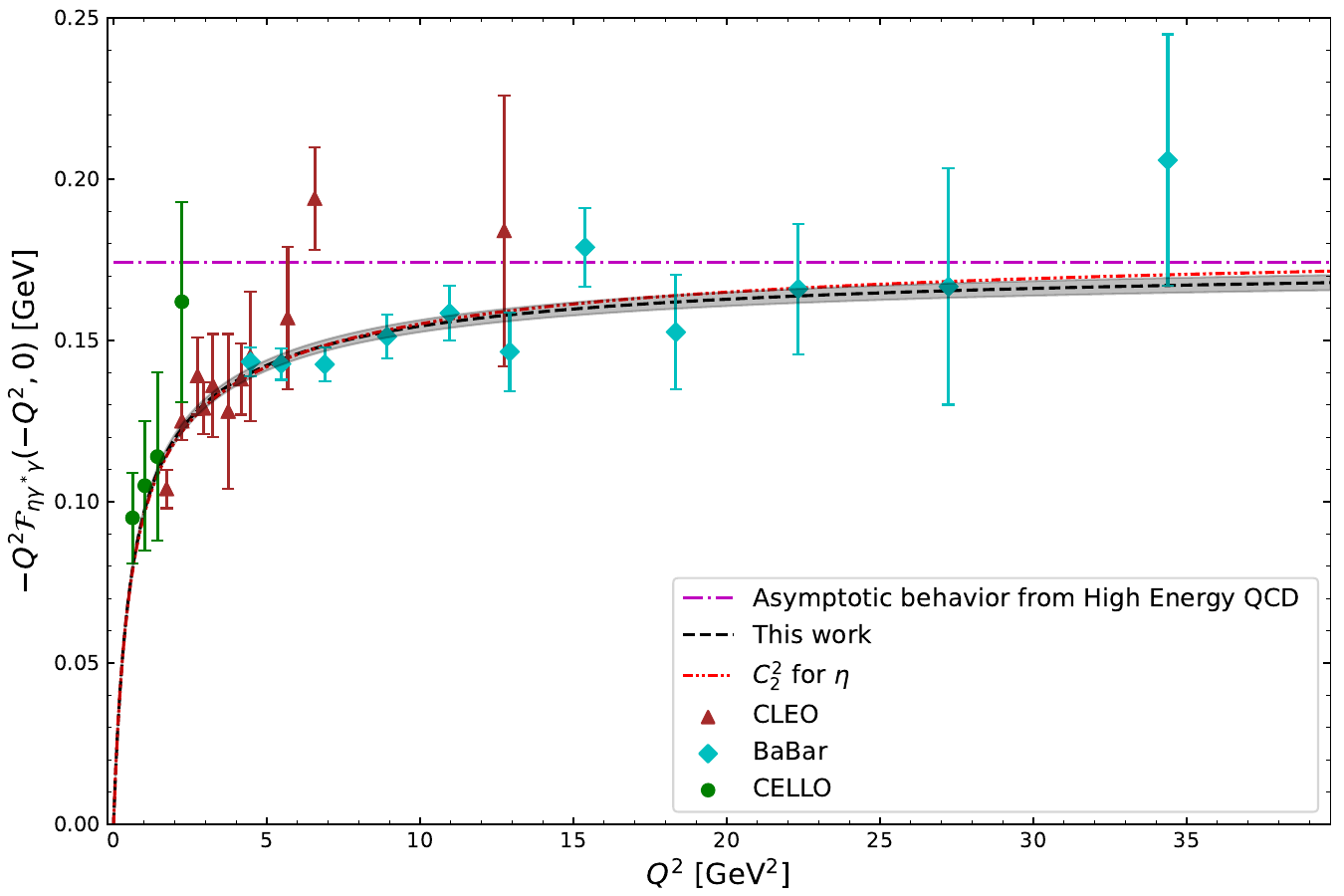}\\[1ex]
\centering\includegraphics[width=0.73\textwidth]{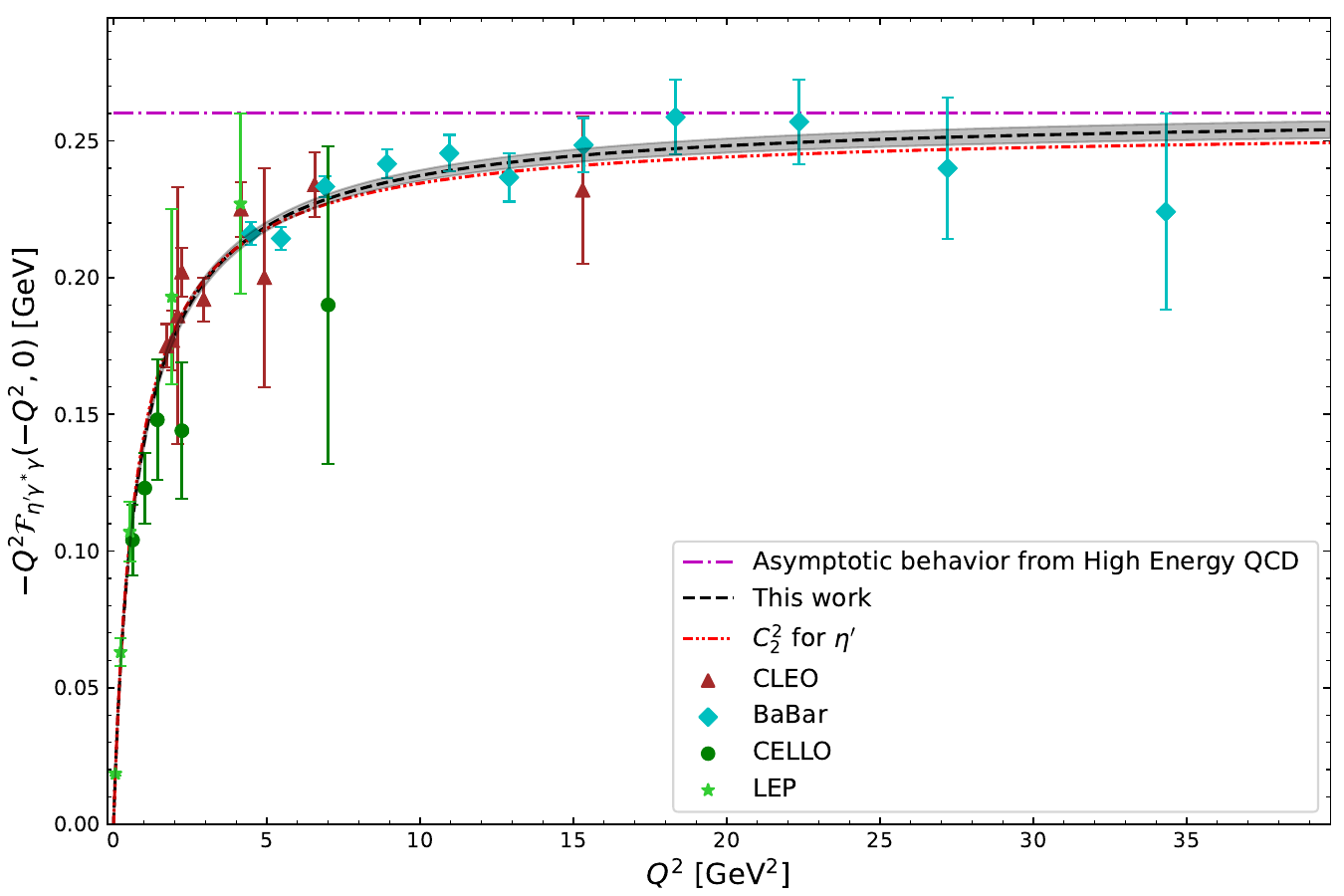}
    \caption{The comparison between our best fit (table~\ref{tab:completefitresults}) for R$\chi$T and the $C_2^2$ results in the singly virtual regime for $\pi^0$, $\eta$ and $\eta^\prime$.}
    \label{fig:THSandC221V}
\end{figure}

\begin{figure}
\centering\includegraphics[width=0.73\textwidth]{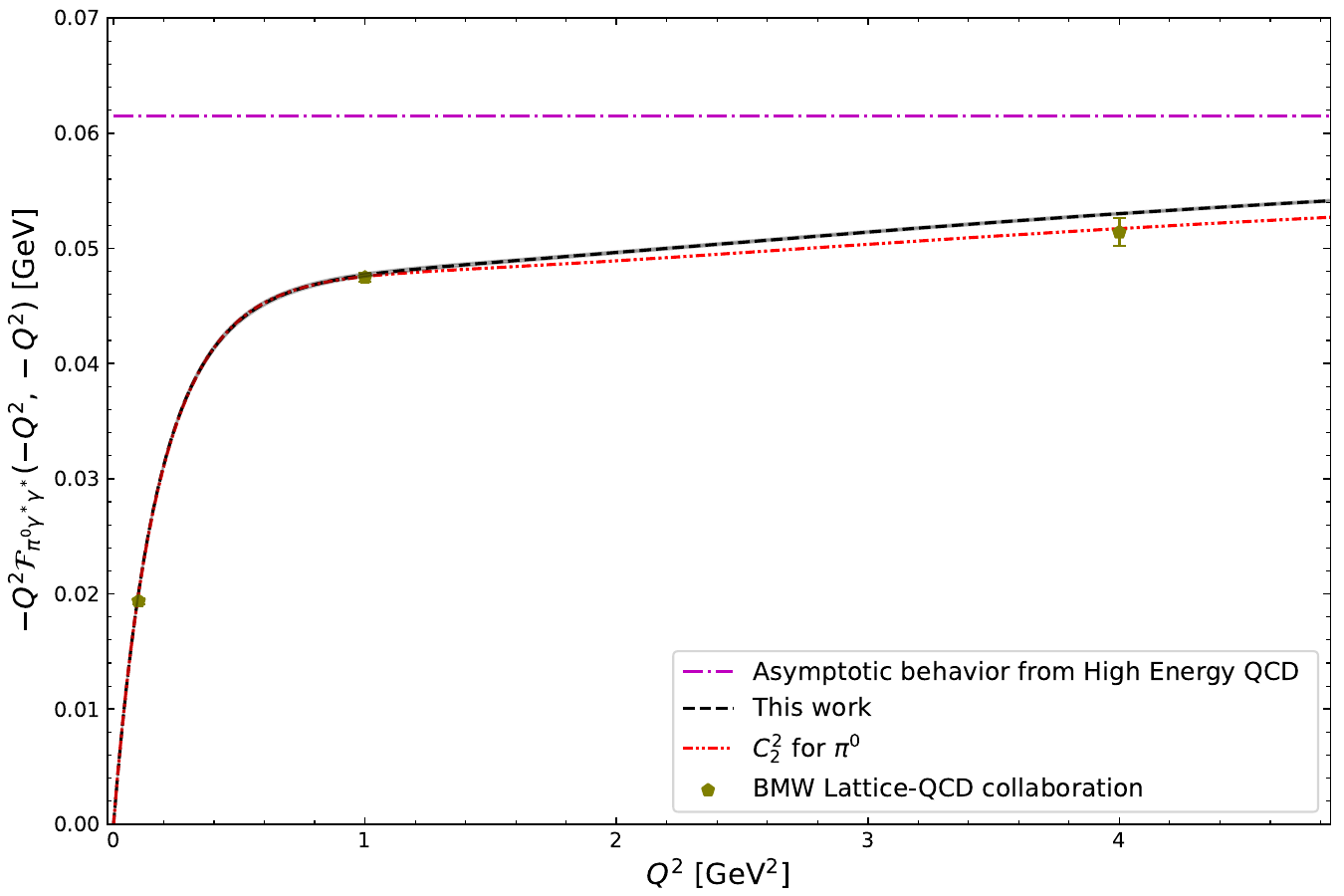}\\[1ex]
\centering\includegraphics[width=0.73\textwidth]{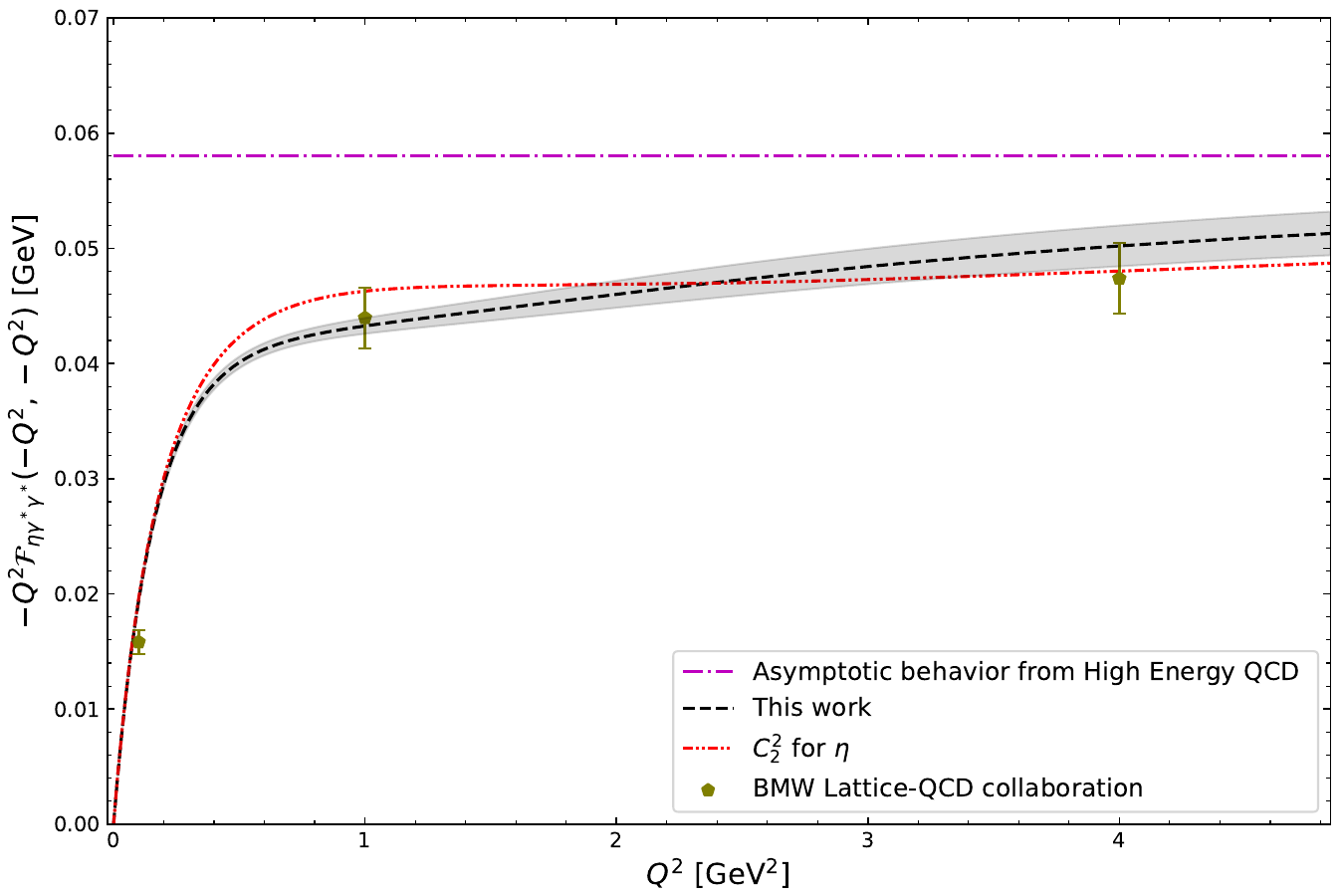}\\[1ex]
\centering\includegraphics[width=0.73\textwidth]{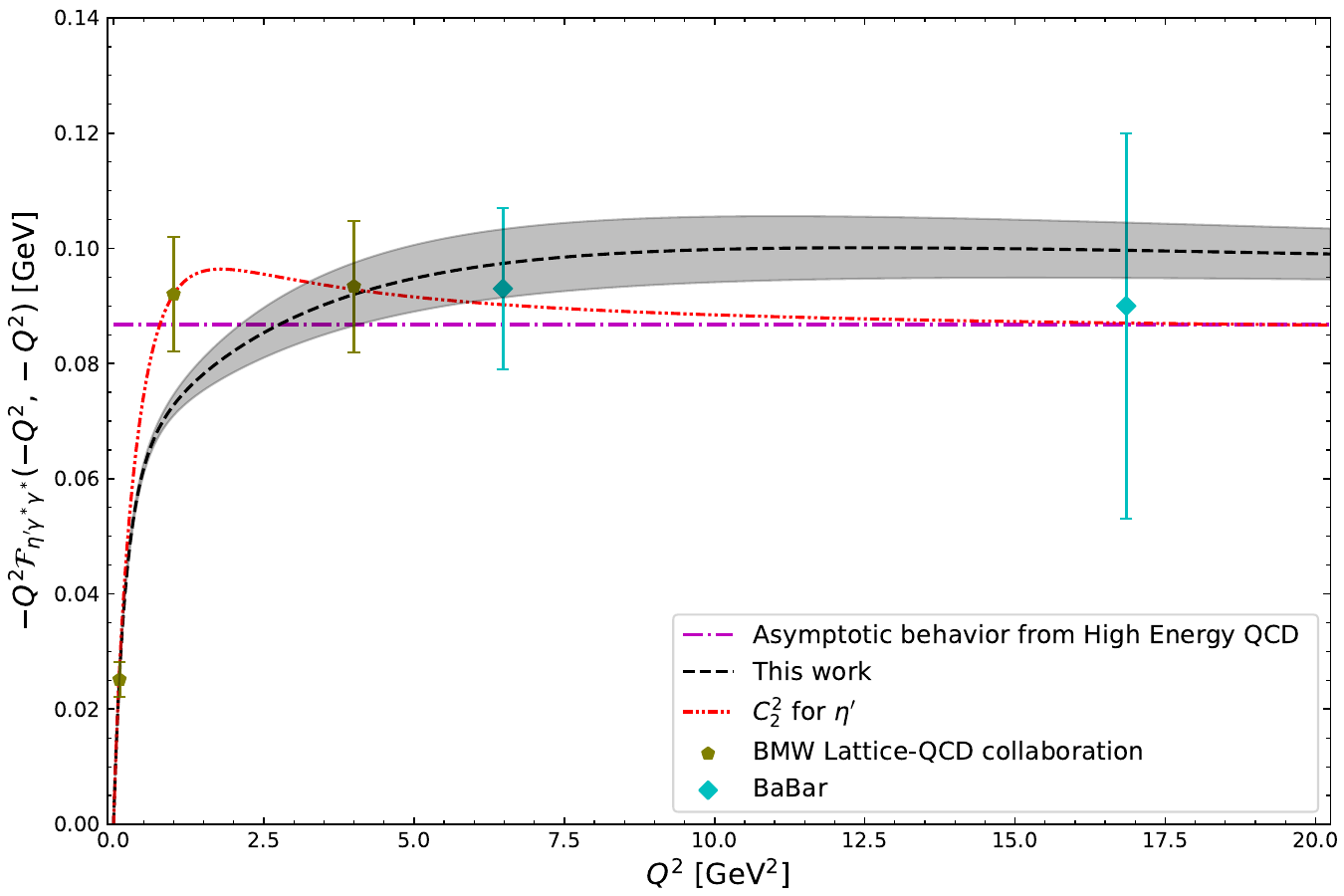}
      \caption{The comparison between our best fit (table \ref{tab:completefitresults}) for R$\chi$T and the $C_2^2$ results in the doubly virtual regime for $\pi^0$, $\eta$ and $\eta^\prime$.}
    \label{fig:THSandC222V}
\end{figure}

Therefore, in this section we will be comparing the following fits to data:
\begin{itemize}
\item Our best fit result, described in sec.~\ref{sec_Fits}, dubbed R$\chi$T below.
\item The results obtained considering the chiral limit of eqs.~(\ref{piTFFfinal}) and (\ref{etaTFFfinal}), which correspond to a $C_2^2$ CA, as indicated in table~\ref{tab:ChiralLimitTHStoCA}. 
This is christened $\chi$R$\chi$T in what follows.
\item The update of the results in ref.~\cite{Masjuan:2017tvw}, using a $C_2^1$. 
In this case we change their procedure for the $\eta^{(\prime)}$ cases, as explained in the previous paragraph.
\item The fit with a $C_2^2$~\footnote{This was performed as a minimal extension of the work in~\cite{Masjuan:2017tvw}, with the same procedure of the update previously described. As mentioned before, this model has 3 free parameters sensitive exclusively to double virtuality data.} demanding that no poles are generated in the unphysical region. This has also been done  using the results in table~\ref{tab:ChiralLimitTHStoCA} to determine the $C_2^2$.
\end{itemize}
The results of these fits are shown in table~\ref{tab:alphachisqd}. 
According to them, the best agreement with data is obtained for R$\chi$T, and only $C_2^2$ 
yields fits with a reduced $\chi^2/{\rm{dof}}\sim1$ as well. 
Figures~\ref{fig:THSandC221V} and \ref{fig:THSandC222V} extend Figs.~\ref{fig:TFF1V} and \ref{fig:TFF2V} by adding the $C_2^2$ results. In the doubly virtual case, Fig.~\ref{fig:THSandC222V}, it is clear that there are very few points to fit the three coefficients that are independent for each pseudoscalar in $C_2^2$, causing all 3 pseudoscalar mesons to have the parameter $\beta^P_{22}$ at limit; consequently, there are poles within the 1$\sigma$ region of these parameterizations of the TFFs. 
On the contrary, the 3 couplings sensitive to double virtuality in R$\chi$T are related by flavor symmetry, so they are fitted to 14 data points, which prevents overfitting.
\begin{table}[h!]
    \centering
    \begin{tabular}{| c c c c c | 
    }
    \hline
        Particle  
        & $\chi^2_{\rm{R\chi T}}$/dof & $\chi^2_{\rm{\chi R\chi T}}$/dof & $\chi^2_{C_2^1}$/dof  & $\chi^2_{C_2^2}$/dof \\
        \hline
        $\pi^0$ 
        & 33.3/39 & 58.2/40 & 234.0/40  & 35.18/38 \\
        $\eta$ 
        & 47.7/27 & 61.6/29 & 63.0/31  & 44.9/29 \\
        $\eta^\prime$ 
        & 50.3/36 & 208.5/38 & 42.39/40 & 33.6/38\\
        \hline
    \end{tabular}
    \caption{Our best fit (section~\ref{sec_Fits}) and its chiral limit are compared to the results obtained using CA. In the $C_2^1$, case we obtained $a_{P;1,1}=0.0048(9),0.75(13),2.677(25)$ for $P=\pi^0,\eta,\eta^\prime$, respectively.}
    \label{tab:alphachisqd}
\end{table}

\section{$\pi^{0},\eta,\eta^{\prime}$-pole contributions to $a_\mu$}\label{sec_amuPpole}

In this section we quote the results of our evaluation of the $\pi^0/\eta/\eta^\prime$-pole contributions to the hadronic light-by-light piece of the muon anomalous magnetic moment, $a_\mu^{P\text{-}\rm{pole,\,HLbL}}$.\footnote{In order to lighten the notation, we shall suppress from this point forward the superscript HLbL in $a_{\mu}$, {\it{i.e.}}, we will take $a_\mu^{P\text{-}\rm{pole,\,HLbL}}\rightarrow a_\mu^{P\text{-}\rm{pole}}$.}
We have computed the $a_\mu^{P\text{-}\rm{pole}}$ contributions according to refs.~\cite{Nyffeler:2016gnb}:
\begin{equation}
    a_\mu^{P\text{-}\rm{pole}}=-\frac{2\alpha^3}{3\pi^2}\int_0^\infty dQ_1 dQ_2 \int_{-1}^1 dt \sqrt{1-t^2}Q_1^3 Q_2^3 \left[F_1 P_6 I_1(Q_1,Q_2,t)+F_2 P_7 I_2(Q_1,Q_2,t)\right]\,,
    \label{amu}
\end{equation}
where $\alpha$ is the fine structure constant, $Q_i=|Q_i|$, $t=\cos\theta$, $P_6=\frac{1}{Q_2^2+m_\pi^2}$, $P_7=\frac{1}{Q_3^2+m_\pi^2}$, $Q_3^2=Q_1^2+Q_2^2+2Q_1 Q_2 t$ and the $I_{1(2)}(Q_1,Q_2,t)$ are given in \cite{Nyffeler:2016gnb}.
The information of the transition form factors is encoded in:
\begin{subequations}
\begin{equation}
    F_1=\mathcal{F}_{P\gamma^*\gamma^*}(Q_1^2,Q_3^2)\mathcal{F}_{P\gamma^*\gamma}(Q_2^2,0)\,.
\end{equation}
\begin{equation}
    F_2=\mathcal{F}_{P\gamma^*\gamma^*}(Q_1^2,Q_2^2)\mathcal{F}_{P\gamma^*\gamma}(Q_3^2,0)\,.  
\end{equation}
\end{subequations}

In table \ref{tab:amuEvaluations} we collect different recent evaluations of $a_\mu^{P\text{-}\mathrm{poles}}$, including ours.\footnote{Very recently, the RBC/UKQCD Lattice collaboration has presented a preliminary value for the $\pi^{0}$-pole, $a_{\mu}^{\pi^{0}\text{-}\rm{pole}}=57.8(1.9)_{\rm{stat}}(1.0)_{\rm{syst}}\times10^{-11}$~\cite{RBCUKQCD}, while the values of $a_{\mu}^{\eta\text{-}\rm{pole}}=14.64(77)\times10^{-11}$ and $a_{\mu}^{\eta^{\prime}\text{-}\rm{pole}}=13.44(70)\times10^{-11}$~\cite{EtaEtaPDispersive} have been reported for the $\eta/\eta^{\prime}$-poles in a dispersive approach.} We implemented a numerical evaluation, using the VEGAS algorithm \cite{Lepage:2020tgj,Lepage:1977sw}. For the computation of the statistical error, we generated 1000 sets of points in the parameter space from a normal multivariate distribution, given the central values and correlations from tables \ref{tab:completefitresults} and \ref{tab:fitcorrelations}, which were used for the determination of the mean and standard deviation of each $a_\mu^{\mathrm{P-pole}}$.
\begin{table}[h!]
    \centering
    \begin{tabular}{| l  l  l  l l |} \hline
$\pi^0$ & $\eta$ & $\eta^\prime$ & Ref. & Method\\
\hline
$63.0^{+2.7}_{-2.1}$& - & - & \cite{Hoferichter:2018dmo,Hoferichter:2018kwz}& Dispersive\\
$63.6\pm2.7$& $16.3\pm1.4$ & $14.5\pm1.9$& \cite{Masjuan:2017tvw} & CA\\
$62.6\pm1.3$& $15.8\pm1.2$ & $13.3\pm0.9$ & \cite{Eichmann:2019tjk} & Dyson-Schwinger eqs.\\
$61.4\pm2.1$& $14.7\pm1.9$ & $13.6\pm0.8$ & \cite{Raya:2019dnh}& Dyson-Schwinger eqs.\\
$63.0^{+2.7}_{-2.1}$& $16.3\pm1.4$ & $14.5\pm1.9$&~\cite{Aoyama:2020ynm} & WP Data-driven\\
$62.3\pm2.3$ & - & - & \cite{Aoyama:2020ynm,Gerardin:2019vio} & WP Lattice\\
$57.8\pm2.0$ & $11.6\pm2.0$ & $15.7\pm4.3$ & \cite{Gerardin:2023naa} & Lattice\\
$61.7\pm2.0$ & $13.8\pm5.5$ & - & \cite{ ExtendedTwistedMass:2023hin,ExtendedTwistedMass:2022ofm} & Lattice\\
$63.5\pm0.8$ & - & - & \cite{Kadavy:2022scu} & R$\chi$T, $3$ resonance multiplets\\
$61.9\pm0.6$ & $15.2\pm0.5$ & $14.2\pm0.7$ & This work & R$\chi$T, $2$ resonance multiplets\\
\hline
    \end{tabular}
    \caption{Different recent evaluations of $a_\mu^{P\text{-}\rm{poles}},\,P=\pi^0,\,\eta,\,\eta^\prime$, multiplied by $10^{11}$. 
    In the first rows we collect the results quoted in the WP~\cite{Aoyama:2020ynm}. 
    The last rows include later results. 
    Our uncertainties are only statistical. Our systematic theory errors are assessed in the remainder of this section.}     \label{tab:amuEvaluations}
\end{table}

We summarize the most important differences between our treatment and the other recent R$\chi$T analysis~\cite{Kadavy:2022scu} in table~\ref{tab:DiffswithPrague}. 
The approaches followed in earlier studies~\cite{Kampf:2011ty,Roig:2014uja,Guevara:2018rhj} were already introduced in section \ref{sec_Intro}.
\begin{table}[h!]
    \centering
    \begin{tabular}{| l  l |} \hline
Ref.~\cite{Kadavy:2022scu} & This work\\
\hline
Ansätze & Computation from the R$\chi$T Lagrangian\\
3 resonance multiplets & 2 resonance multiplets\\
Chiral Limit & Chiral symmetry breaking 
up to $\mathcal{O}(m_P^2)$\\
Only $\pi^0$ (for which the chiral limit works quite well) & $\pi^0,\eta,\eta^\prime$\\
More subleading OPE constraints & Only $\delta_\pi$ as subleading OPE constraint\\
2019 Lattice data~\cite{Gerardin:2019vio} & 2023 Lattice data~\cite{Gerardin:2023naa}\\
Only data used & BaBar $\pi^0$ data not used\\
\hline
    \end{tabular}
    \caption{Summary of the main differences between our study and ref.~\cite{Kadavy:2022scu}.}
    \label{tab:DiffswithPrague}
\end{table}

\subsection{Assessment of systematic theory uncertainties}\label{sec_TheoryErrors}

In the following, we will evaluate here the error on $a_\mu^{P\text{-}\rm{poles}}$ associated to our R$\chi$T description of the $P$-TFF. 
Specifically, we will consider the one stemming from the use of different data sets (whether including/excluding BaBar $\pi^0$ single virtual data in the global fit), the one coming from cutting the infinite tower of $V$ and $P$ states to two multiplets, the one arising from neglecting subleading corrections in the large-$N_C$ expansion and, finally, the one coming from excluding the Lattice QCD data for the doubly virtual TFF in our fits. 
We have verified that other corrections (from e.g. neglecting higher-order terms in $m_P^2$, modifying the $\eta$-$\eta^\prime$ mixing parameters according to the NNLO $U(3)$ $\chi$PT fit to lattice data of ref.~\cite{Guo:2015xva}, etc.) are negligible with respect to these (see also the related discussion in ref.~\cite{Guevara:2018rhj}, our relative errors on them are very similar to those reported therein).\footnote{Also negligible are the uncertainties associated to either using $\pi^0$ Lattice data from ref.~\cite{Gerardin:2019vio} or~\cite{Gerardin:2023naa}, and to including or not additional subleading OPE constraints 
(see table \ref{tab:DiffswithPrague}).} Although its associated uncertainty is also negligible, we end this section discussing the asymptotic behaviour for asymmetric double virtualities.

\subsubsection*{Use of all available experimental data}\label{sec_DataSet}

There are several data sets for the $P$-TFF. 
Analyses in the literature differ by including/excluding some experimental data or imposing cuts to the fitted data. 
In this work, as in ref.~\cite{Guevara:2018rhj}, BaBar data for the single virtual region of $\pi^0$ was excluded for the aforementioned reasons in the global fit of the three channels. 
An estimation of the error induced by this decision was computed by comparing the evaluation of the $a_\mu^{P\text{-}\rm{pole}}$ obtained by including or excluding these data. 
The obtained difference for each pseudoscalar meson is:
\begin{subequations}
    \begin{equation}
        \left(\Delta a_\mu^{\pi^{0}\text{-}\rm{pole}}\right)_{\mathrm{Data\,Sets}}\,=+0.20\times10^{-11},
    \end{equation}
    \begin{equation}
        \left(\Delta a_\mu^{\eta\text{-}\rm{pole}}\right)_{\mathrm{Data\,Sets}}=-0.02\times10^{-11}\,,      
    \end{equation}
    \begin{equation}
        \left(\Delta a_\mu^{\eta^{\prime}\text{-}\rm{pole}}\right)_{\mathrm{Data\,Sets}}=+0.02\times10^{-11}\,.
    \end{equation}
    \label{eq_DataSets}
\end{subequations}
As expected, the induced error occurs mainly in the evaluation of $a_\mu^{\pi^{0}\text{-}\rm{pole}}$; however, small deviations are present for $\eta$ and $\eta^\prime$ because R$\chi$T connects the TFFs of the 3 particles trough chiral symmetry.

\subsubsection*{Finite number of resonances}\label{sec_Cuttingtower}

Even though the full R$\chi$T has an infinite tower of resonances, cutting to a finite number is required for practical use (unless there exists an exact resummation mechanism, which usually only happens for the simplest toy models). 
This analysis includes -besides the pseudo-Goldstone bosons $P$- two resonance multiplet states for the vector mesons $V$, $V^\prime$, and one for the pseudoscalar mesons $P^\prime$. 
A minimal extension to this model was performed earlier in ref.~\cite{Kadavy:2022scu} -referred as three-multiplet resonance-, where a third $V^{\prime \prime}$ and $P^{\prime\prime}$ were included, albeit in the chiral limit, and only for the $\pi^0$ meson.

An estimation of the systematic error caused by having a finite number of resonances was performed using the results from this three-multiplet resonance model and a fit to the data of our model in the chiral limit, $\chi$R$\chi$T. 
In ref.~\cite{Kadavy:2022scu}, the value of $a_\mu^{\rm{\pi^{0}\text{-}pole}}$ was computed using the $\pi^0$ lattice data from ref.~\cite{Gerardin:2019vio} up to $Q^2=4 \mathrm{GeV}^2$ to fit their free parameters, so an equivalent analysis (with the same dataset) was performed in $\chi$R$\chi$T to fit the 3 free parameters of table~\ref{tab:ChiralLimitTHStoCA} -with the mixing parameters fixed to the results of our best fit-- using only the $\pi^0$ lattice data points from ref.~\cite{Gerardin:2023naa}. 
For $\eta$ and $\eta^\prime$, there is no computation of the pole contributions to the HLbL piece of $a_\mu$, so we cast the overall $c_P$ flavor-space-rotation factor (defined just above table~\ref{tab:ChiralLimitTHStoCA}) to compute a fair estimate on these contributions within the three-multiplet resonance model. The resulting differences for each particle are:
\begin{subequations}
    \begin{equation}
        \left(\Delta a_\mu^{\pi^0\text{-}\rm{pole}}\right)_{\mathrm{finite \, spectrum}}=+1.8\times10^{-11}\,,
    \end{equation}
    \begin{equation}
        \left(\Delta a_\mu^{\eta\text{-}\rm{pole}}\right)_{\mathrm{finite \, spectrum}}=+1.0\times 10^{-11}\,,
    \end{equation}
    \begin{equation}
        \left(\Delta a_\mu^{\eta^{\prime}\text{-}\rm{pole}}\right)_{\mathrm{finite \, spectrum}}=+1.4\times 10^{-11}\,.
    \end{equation}   
    \label{eq_finitespectrum}
\end{subequations}

\subsubsection*{Subleading corrections in $1/N_C$}\label{sec_NLO1/N}

We have considered the contributions at leading order in the large-$N_C$ expansion in our computation. We have estimated the impact of neglected higher-order effects by considering the modification to the $\rho$ propagator coming from pion and kaon loops at next-to-leading order (NLO).\footnote{We have computed this modification up to $1$ GeV. At higher energies other effects arise (like, e.g. those associated to inelasticities), with a relevant interplay to this one, need to be accounted for as well.} 
Although a proper analysis of the SD behavior for the TFF should be done, we will use the result from the electromagnetic form factor of the pion \cite{Guerrero:1997ku}, which will amount to the following replacement (the $\rho$ propagator remains real in the whole space-like region, $q^2<0$)
\begin{equation}
M_\rho^2-q^2\to M_\rho^2-q^2+ \frac{q^2M_\rho^2}{96\pi^2F_\pi^2}\left(A_\pi (q^2) + \frac{1}{2} A_K (q^2)\right)\,,
\end{equation}
where
\begin{equation}
A_P(q^2)=\ln\frac{m_P^2}{M_\rho^2}+8\frac{m_P^2}{q^2}-\frac{5}{3}+\sigma_P^3(q^2)\ln\left(\frac{\sigma_P(q^2)+1}{\sigma_P(q^2)-1}\right)\,,
\end{equation}
with $\sigma_P(q^2)=\sqrt{1-\frac{4m_P^2}{q^2}}$. 
We will take its absolute value (there are other types of corrections at this order that we are disregarding, like those to the $VVP$ vertex, that can have either sign) as the one standard deviation uncertainty induced by missing subleading corrections in the large-$N_C$ limit. 
Then, we will have: 
\begin{subequations}
\label{eq_error3}
\begin{equation}
\left(\Delta a_\mu^{\pi^0\text{-}\rm{pole}}\right)_{1/N_C}=\pm1.5\times10^{-11}\,,
\end{equation}
\begin{equation}
\left(\Delta a_\mu^{\eta\text{-}\rm{pole}}\right)_{1/N_C}=\pm0.5\times10^{-11}\,,
\end{equation}
\begin{equation}
\left(\Delta a_\mu^{\eta^{\prime}\text{-}\rm{pole}}\right)_{1/N_C}=\pm0.3\times10^{-11}\,.
\end{equation}
\end{subequations}

\subsubsection*{Combination of experimental and Lattice data}

Finally, we evaluate the uncertainty on $a_{\mu}^{P\text{-}\rm{poles}}$ that comes from neglecting Lattice QCD data in our fits.
In this way, we quantify the difference between a fully data-driven analysis and a hybrid one.
The differences, for every $P$, are:
\begin{subequations}\label{eq_latticeError}
    \begin{equation}
\left(\Delta a_\mu^{\pi^0\text{-}\rm{pole}}\right)_{\mathrm{Hybrid\,analysis}}=+0.4\times10^{-11}\,,
    \end{equation}    
    \begin{equation}
\left(\Delta a_\mu^{\eta\text{-}\rm{pole}}\right)_{\mathrm{Hybrid\,analysis}}=-0.6\times10^{-11}\,,
    \end{equation} 
    \begin{equation}
\left(\Delta a_\mu^{\eta^{\prime}\text{-}\rm{pole}}\right)_{\mathrm{Hybrid\,analysis}}=-0.8\times10^{-11}\,.
    \end{equation}     
\end{subequations}
The $\pi^0$-pole increases while the $\eta/\eta^{\prime}$-poles decrease by including the Lattice data.
To the best of our knowledge, this is the first quantification of the difference between a hybrid and a fully data-driven analysis and it is possible here due to chiral symmetry, since the parameters $d_{d1}$, $d_{d2}$ and $d_{d3}$ appear in the $\pi^0,\eta$ and $\eta^\prime$ TFFs -even though the experimental data is available only for $\eta^\prime$, a joint description for all the 3 particles can be obtained. 
This situation contrasts with the case of the Canterbury Approximants analysis presented in table \ref{tab:alphachisqd}, where Lattice data cannot be excluded for $\pi^0$ nor for $\eta$.
\subsubsection*{Asymptotic Behavior for asymmetric double virtualities}
\begin{figure}[h!]
    \centering
    \includegraphics[width=0.8\linewidth]{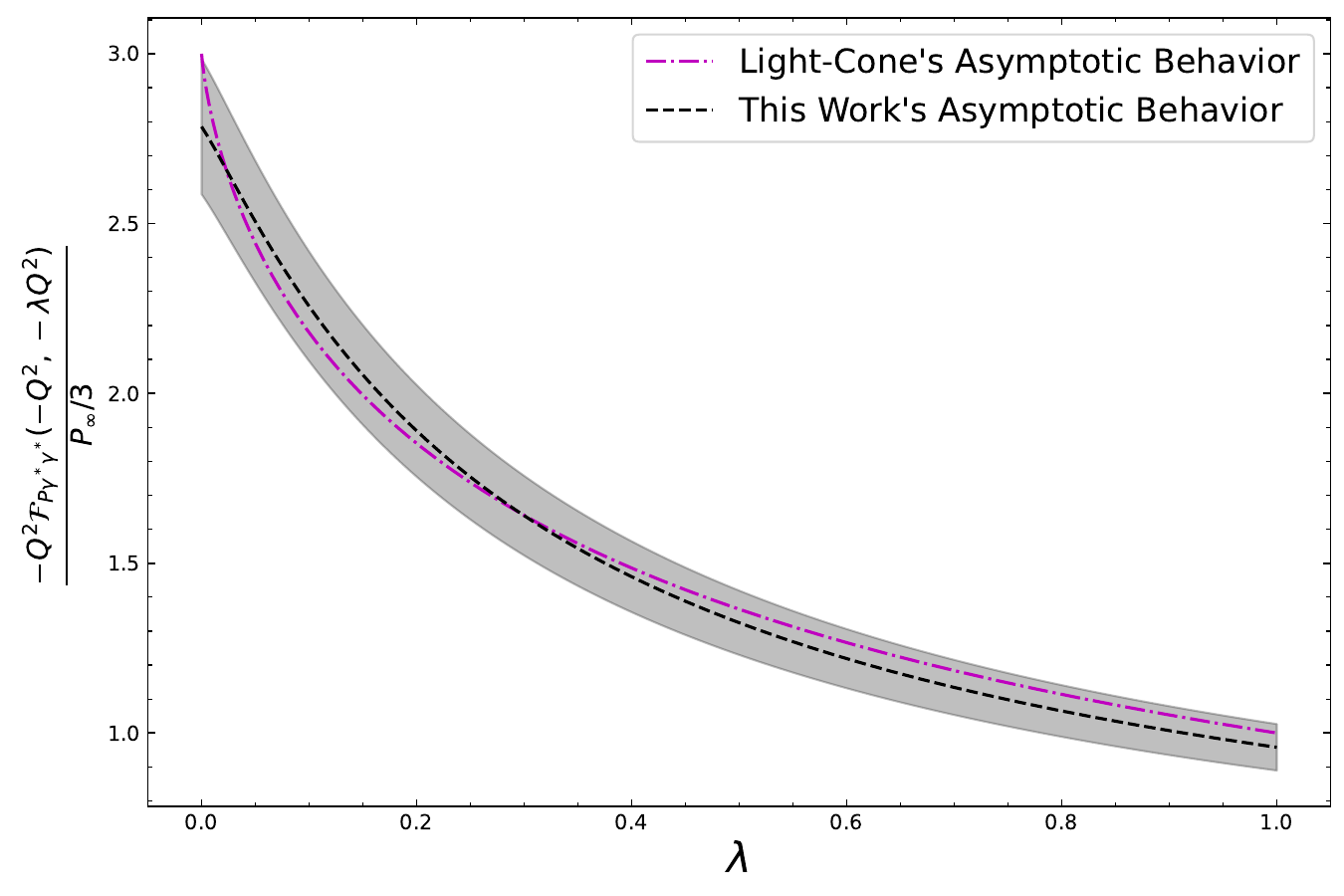}
    \caption{Comparison between the asymptotic behavior of the light-cone expansion and this work's TFFs in terms of the ratio between the 2 squared momenta $\lambda=Q_1^2/Q_2^2$. Our one $\sigma$ uncertainties are represented by the gray band.}
    \label{fig:asymmetricSDC}
\end{figure}

Furthermore, in this work we have imposed the single virtual and symmetric double virtual SDCs, (\ref{eq_SDCspi0}). However, from the light-cone expansion \cite{Lepage:1980fj,Lepage:1979zb,Agaev:2010aq}, the SDCs at leading order in pQCD and at leading-twist are given for general large asymmetric virtualities, by:
\begin{equation}
    \lim_{Q_{1(2)}^2\to\infty} \mathcal{F}_{P\gamma^*\gamma^*}(-Q_1^2,-Q_2^2)=\frac{P_\infty}{3} \int_0^1 dx \frac{\phi_P(x)}{xQ_1^2+(1-x)Q_2^2},
    \label{eq_lightconeSDC}
\end{equation}
where $\phi_P(x)$ is the pion distribution amplitude which -for large momenta- behaves as $\phi_P(x)\to 6x(1-x)$\cite{Lepage:1980fj,Braun:2003rp}. The construction of the TFFs using R$\chi$T is limited to a polynomial description, which cannot reproduce the whole range of asymmetries in eq.~(\ref{eq_lightconeSDC}). The asymmetric SDCs, $\lim_{Q^2\to\infty} \mathcal{F}_{P\gamma^*\gamma^*}(-Q^2,-\lambda Q^2)$, were compared -for $R\chi T$ and the light-cone expansion results- within the relevant region of the integration kernels for the $a_\mu^{\mathrm{P-pole}}$, as it is shown in fig. \ref{fig:asymmetricSDC}. There is a bump in the asymptotic behavior of eqs. (\ref{piTFFfinal}) and (\ref{etaTFFfinal}) because the term reproducing the symmetric double virtual SDC of eq.(\ref{eq_SDCspi0}) dominates the whole scale of asymmetries except for $\lambda\to 0$, and it behaves as $1/\lambda$.

To quantify the effect in the evaluation of $a_\mu^{\mathrm{P-pole}}$ produced by this difference, a replacement of the dominant terms at high energies of eqs.(\ref{piTFFfinal}) and \ref{etaTFFfinal}) by the one of eq. (\ref{eq_lightconeSDC}) -as suggested in eq. (5.13) from \cite{Hoferichter:2018kwz}- was done, and the numerical difference was one order of magnitude smaller than the rest of the theory errors for the 3 pseudoscalar mesons. The change for the expression of the light-cone expansion was performed at a $Q_1$ where the squared difference between this work's asymptotic behavior and eq. (\ref{eq_lightconeSDC}) is minimized.
\subsection{Final results with statistical and systematic errors}

With the four (independent) dominant uncertainties given by eqs.~(\ref{eq_DataSets}), (\ref{eq_finitespectrum}), (\ref{eq_error3}) and (\ref{eq_latticeError}), we obtain the following systematic theory error:
\begin{subequations}
    \begin{equation}
    \left(\Delta a_\mu^{\pi^{0}\text{-}\rm{pole}}\right)_{\rm{theory}}=\left({}^{+2.4}_{-1.5}\right)\times10^{-11}\,,
    \label{pi0theoryerror}
    \end{equation}    
    \begin{equation}
    \left(\Delta a_\mu^{\eta\text{-}\rm{pole}}\right)_{\rm{theory}}=\left({}^{+1.1}_{-0.8}\right)\times10^{-11}\,,
    \label{etatheoryerror}
    \end{equation} 
    \begin{equation}
    \left(\Delta a_\mu^{\eta^{\prime}\text{-}\rm{pole}}\right)_{\rm{theory}}=\left({}^{+1.4}_{-0.9}\right)\times10^{-11}\,.
    \label{etaPtheoryerror}
    \end{equation}     
\end{subequations}
Our final result, including statistical (cf.~table~\ref{tab:amuEvaluations}) and the above systematic uncertainties 
\footnote{We always quote the systematic error after the statistical one.} is:
\begin{subequations}
\begin{equation}
    a_\mu^{\pi^{0}\text{-}\rm{pole}}=\left(61.9\pm0.6^{+2.4}_{-1.5}\right)\times10^{-11}=\left(61.9^{+2.5}_{-1.6}\right)\times10^{-11}\,,
    \label{amupi0final}
\end{equation}
\begin{equation}
    a_\mu^{\eta\text{-}\rm{pole}}=\left(15.2\pm0.5^{+1.1}_{-0.8}\right)\times10^{-11}=\left(15.2^{+1.2}_{-0.9}\right)\times10^{-11}\,,
    \label{amuetafinal}
\end{equation}
\begin{equation}
    a_\mu^{\eta^\prime\text{-}\rm{pole}}=\left(14.2\pm0.7^{+1.4}_{-0.9}\right)\times10^{-11}=\left(14.2^{+1.6}_{-1.1}\right)\times10^{-11}\,,
    \label{amuetaPfinal}
\end{equation}
\end{subequations}
with an uncertainty saturated by the model-dependence. 
Combining eqs.~(\ref{amupi0final}), (\ref{amuetafinal}) and (\ref{amuetaPfinal}) we arrive at the following result for the pseudoscalar-pole contributions:
\begin{equation}\label{eq_mainresult}
a_\mu^{\pi^0+\eta+\eta^{\prime}\text{-}\rm{pole}}=\left(91.3\pm1.0^{+3.0}_{-1.9}\right)\times10^{-11}=\left(91.3^{+3.2}_{-2.1}\right)\times10^{-11}\,.
\end{equation}

\section{Conclusions}\label{sec_Concl}


In this work, we have presented a simultaneous description of the singly and doubly virtual $\pi^{0},\eta$ and $\eta^{\prime}$ transition form factors based on Resonance Chiral Theory that complies with the White Paper consensus agreement for taking their $a_\mu^{\rm{HLbL}}$ contribution into account.
In particular, we have shown that working within the two resonance multiplets saturation scheme we: satisfy leading (and some subleading) chiral and high-energy QCD constraints; get a normalization given by the two-photon partial decay width, $\Gamma(\pi^{0}/\eta/\eta^{\prime}\to\gamma\gamma)$, that is fully compatible with experimental values; reproduce the singly virtual transition form factors experimental data in the spacelike region (see Fig.~\ref{fig:TFF1V}) and, last but not least; obtain a faithful description of the doubly virtual transition form factor for all three pseudoscalar mesons resulting from the use of the BaBar data in the $\eta^{\prime}$ channel in combination with Lattice-QCD results for the three mesons form factors (see Fig.~\ref{fig:TFF2V}).
Our evaluation of the pole contributions to the hadronic light-by-light piece of the muon $g-2$ read:
$a_\mu^{\pi^{0}\text{-}\rm{pole}}=\left(61.9\pm0.6^{+2.4}_{-1.5}\right)\times10^{-11}$, $a_\mu^{\eta\text{-}\mathrm{pole}}=\left(15.2\pm0.5^{+1.1}_{-0.8}\right)\times10^{-11}$ and $a_\mu^{\eta^\prime\text{-}\rm{pole}}=\left(14.2\pm0.7^{+1.4}_{-0.9}\right)\times10^{-11}$, for a total of $a_\mu^{\pi^0+\eta+\eta^{\prime}\text{-}\rm{pole}}=\left(91.3\pm1.0^{+3.0}_{-1.9}\right)\times10^{-11}$, where the first error is statistical and the second one is systematic (see Sec.~\ref{sec_TheoryErrors}).

Our determination for $a_\mu^{\pi^{0}\text{-}\rm{pole}}$ fulfills the quality criteria outlined above being compatible at the level of $1\sigma$ with the dispersive evaluation, $a_\mu^{\pi^{0}\text{-}\rm{pole}}=\left(63.0^{+2.7}_{-2.1}\right)\times10^{-11}$~\cite{Aoyama:2020ynm} and with the results based on Canterbury approximants, $a_\mu^{\pi^{0}\text{-}\rm{pole}}=63.6(2.7)\times10^{-11}$~\cite{Masjuan:2017tvw}, and with the most recent lattice QCD results, $a_\mu^{\pi^{0}\text{-}\rm{pole}}=57.8(2.0)\times10^{-11}$ \cite{Gerardin:2023naa},~\footnote{This result and our work agree within 1.6 $\sigma$, which is the biggest among the different works considered, but still reasonable.} and  $a_\mu^{\pi^{0}\text{-}\rm{pole}}=61.7(2.0)\times10^{-11}$\cite{ExtendedTwistedMass:2023hin}.
Our outcomes for the $a_\mu^{\eta/\eta^\prime\text{-}\rm{pole}}$ contributions are particularly relevant, given that our approach complies with the leading asymptotic behavior for double virtuality and the good performance exhibited for describing the best doubly-virtual input at disposal. 
Interestingly enough, they are consistent with $a_\mu^{\eta/\eta^\prime\text{-}\rm{pole}}=[16.3(1.4)/14.5(1.9)]\times10^{-11}$ from Canterbury Approximants~\cite{Masjuan:2017tvw}, which however would need a $C_4^4$ to fully account for chiral symmetry constraints and their explicit breaking at leading order, that is non-negligible for the $\eta^{(\prime)}$ contributions (see section \ref{sec_CA}).\footnote{At least a $C_2^2$ would be needed to describe the double virtuality data, instead of the $C_2^1$ used in ref.~\cite{Masjuan:2017tvw}.}
Our results are also compatible with the Lattice results $a_\mu^{\eta/\eta^\prime\text{-}\rm{pole}}=[11.6(2.0)/15.7(4.2)]\times10^{-11}$~\cite{Gerardin:2023naa}, although the $\eta$-pole contribution is in slight tension with our outcome due to the lower value for the normalization of their transition form factor.

We hope that our analysis strengthens the case for experimental measurements of the transition form factors of double virtuality for all three pseudoscalar mesons, as they would allow constraining their functional form and reduce the uncertainty in $a_{\mu}^{P\text{-}\rm{pole}}$.

\acknowledgments
 The authors are grateful to S.~S.~Agaev and P.~Sánchez-Puertas for useful correspondence and advice and are indebted to the anonymous referee for an insightful report that guided us to polish our manuscript. 
E.~J.~E, A.~G. and P.~R. thank partial funding from Conahcyt (México). 
E.~J.~E. also thanks the support of the Inter-American Network of Networks for QCD challenges and of Los Alamos National Laboratory for hospitality -special thanks to E.~Mereghetti and S.~G-S. for hosting his visit -where the initial stages of this work were developed. 
The work of S.~G-S.~is supported by MICIU/AEI/10.13039/501100011033 and 
by FEDER UE through grants PID2023-147112NB-C21; and 
through the ``Unit of Excellence Mar\'ia de Maeztu 2020-202'' 
award to the Institute of Cosmos Sciences, grant CEX2019-000918-M. 
Additional support is provided by the Generalitat de Catalunya (AGAUR) 
through grant 2021SGR01095. 
S.~G-S.~is a Serra H\'{u}nter Fellow. 
P.~R. acknowledges Conahcyt (México) funding through project CBF2023-2024-3226 as well as Spanish support during his sabbatical through projects MCIN/AEI/10.13039/501100011033, grants PID2020-114473GB-I00 and PID2023-146220NB-I00,
and Generalitat Valenciana grant PROMETEO/2021/071.


\end{document}